\documentclass[twocolumn,superscriptaddress,amsmath,amssymb,aps,prc]{revtex4-2}
\usepackage{amsmath,amssymb}
\usepackage{graphicx}
\usepackage{dcolumn}
\usepackage{bm}
\usepackage{multirow}
\usepackage{textcomp}
\usepackage{url}
\usepackage{enumitem}
\usepackage{float}
\usepackage[dvipsnames]{xcolor}
\usepackage[colorlinks=true, allcolors=blue!80!black!]{hyperref}
\usepackage{orcidlink}

\newcommand{\ket}[1]{|{#1}\rangle}
\newcommand{\braket}[2]{\langle{#1}|{#2}\rangle}

\begin{document}

\title{Full configuration interaction quantum Monte Carlo for accurate \textit{ab initio} nuclear structure calculations: algorithms and calculation details}

\author{R. Z. Hu\,\orcidlink{0009-0002-8797-6622}}
\affiliation{School of Physics, and State Key Laboratory of Nuclear Physics and Technology, Peking University, Beijing 100871, China}
\author{F. R. Xu\,\orcidlink{0000-0001-6699-0965}}\email[]{frxu@pku.edu.cn}
\affiliation{School of Physics, and State Key Laboratory of Nuclear Physics and Technology, Peking University, Beijing 100871, China}
\affiliation{Southern Center for Nuclear-Science Theory (SCNT), Institute of Modern Physics, Chinese Academy of Sciences, Huizhou 516000, China}
\author{B. S. Hu\,\orcidlink{0000-0001-8071-158X}}\email[]{baishanhu@pku.edu.cn}
\affiliation{School of Physics, and State Key Laboratory of Nuclear Physics and Technology, Peking University, Beijing 100871, China}
\affiliation{Southern Center for Nuclear-Science Theory (SCNT), Institute of Modern Physics, Chinese Academy of Sciences, Huizhou 516000, China}
\author{A. Alavi\,\orcidlink{0000-0002-0654-9489}}
\affiliation{Max Planck Institute for Solid State Research, Heisenbergstr. 1, 70569 Stuttgart, Germany}
\affiliation{Department of Chemistry, University of Cambridge, Lensfield Road, Cambridge CB2 1EW, United Kingdom}

\date{\today}

\begin{abstract}
    Full configuration interaction quantum Monte Carlo (FCIQMC) is a stochastic many-body solver that has been widely applied to electronic, molecular, and condensed-matter systems. In this work we apply FCIQMC to \textit{ab initio} nuclear structure calculations using interactions derived from chiral effective field theory. We describe the algorithm in detail, including imaginary-time propagation, excitation generation, estimator choices, the initiator approximation with adaptive shift correction, and reduced-density-matrix (RDM) sampling. Benchmark calculations in small model spaces, where deterministic full configuration interaction (FCI) results are available, validate the stochastic calculation of energies, radii, and RDM-based pure estimators. For large model spaces, we analyze the residual finite-walker bias through systematic walker-number convergence and infinite-walker extrapolations. We also demonstrate that FCIQMC can be extended beyond ground-state calculations by computing the low-lying spectrum of $^6$Li.
\end{abstract}

\maketitle

\section{Introduction}
\textit{Ab initio} nuclear theory aims to quantitatively describe finite nuclei and nuclear matter from microscopic nuclear Hamiltonians in a systematically improvable way~\cite{Machleidt2023,Ekstrm2023}. Based on chiral effective field theory~\cite{Epelbaum2009,Machleidt2011,Hammer2020,Tews2022}, nuclear interactions and electroweak currents can be derived order by order~\cite{PhysRevC.78.064002,PhysRevLett.103.102502,PhysRevC.80.034004,PhysRevC.84.024001,PhysRevLett.107.062501,Gysbers2019,PhysRevC.102.025501,Krebs2020,PhysRevLett.132.232503,PhysRevLett.132.232504}. With these developments, the accuracy of the many-body solver itself has become a central issue~\cite{Hergert2020}. The no-core shell model (NCSM)~\cite{PhysRevLett.84.5728,Barrett2013} provides direct access to the full configuration-interaction (FCI) solution, but its reach is limited by the rapidly-growing Hilbert space~\cite{Jiang2026}. Many-body expansion methods, such as many-body perturbation theory (MBPT)~\cite{ROTH2010272,PhysRevC.94.014303,TICHAI2018448,PhysRevLett.122.042501,10.3389/fphy.2020.00164,drischler2026,q3vn-8y8s}, in-medium similarity renormalization group (IMSRG)~\cite{PhysRevLett.106.222502,PhysRevC.85.061304,Hergert2016,PhysRevC.92.034331,PhysRevC.105.L061303,PhysRevC.99.061302,PhysRevLett.103.082501,Zhen2025}, coupled-cluster theory (CC)~\cite{PhysRevLett.113.142502,PhysRevC.76.034302,RevModPhys.79.291,PhysRevLett.92.132501,PhysRevC.82.034330,PhysRevC.83.054306,Hagen2014,p297-y8vq} and self-consistent Green's function (SCGF)~\cite{Dickhoff2004,Som2020,PhysRevC.101.014318}, extend calculations to heavier nuclei, but rely on truncations whose residual many-body uncertainties are difficult to quantify~\cite{PhysRevC.103.044318,PhysRevC.111.034311,PhysRevC.110.044316,PhysRevC.110.044317,PhysRevLett.134.182502,drischler2026,gjmd-fyjy,q3vn-8y8s}. These limitations motivate the development of full-space stochastic solvers that can approach the FCI limit without explicitly storing the many-body Hamiltonian matrix.

Full configuration interaction quantum Monte Carlo (FCIQMC) was first introduced in quantum chemistry as a projector quantum Monte Carlo method in the space of Slater determinants (SDs)~\cite{Booth2009,Booth2010,Booth2011}. Since then it has been widely applied to molecular, solid-state and homogeneous-electron-gas systems~\cite{Shepherd2012,PhysRevB.85.081103,Booth2012n}. In FCIQMC, the configuration interaction (CI) coefficients are represented by a population of signed walkers, which are propagated in imaginary time through stochastic spawning and death/cloning steps. The signed walker population on each determinant provides a stochastic estimate of the corresponding CI coefficient. A central feature of the algorithm is annihilation: walkers of opposite signs on the same determinant are canceled, allowing the fermionic sign structure to be established stochastically~\cite{Booth2009,Booth2010,Booth2011}. The initiator approximation~\cite{Cleland2011} made calculations in much larger Hilbert spaces feasible by restricting spawning from weakly occupied determinants, at the cost of a finite-walker bias that is systematically reduced with increasing walker population. Subsequent developments include reduced-density-matrix sampling for pure estimators and other properties~\cite{Overy2014,Blunt2017}, orthogonalized multi-state propagation for excited states~\cite{Blunt2015_2}, and highly parallel implementations for large-scale calculations~\cite{Booth2014,Guther2020}. Previous studies have placed FCIQMC among the most accurate quantum many-body solvers for realistic electronic Hamiltonians, especially in strongly correlated regimes~\cite{PhysRevX.10.011041,Spencer2019,Dobrautz2025,Dobrautz2019}.

These features make FCIQMC particularly attractive for \textit{ab initio} nuclear structure calculations~\cite{Jin2025}. Nuclear wave functions can contain strong high-order correlations arising from deformation~\cite{PhysRevC.105.L061303,sun2025}, clustering~\cite{PhysRevLett.109.252501,PhysRevLett.112.102501,Otsuka2022}, open-shell structure~\cite{Tichai2024,Tichai2018_2,mmy4-3wrp}, or collective excitations~\cite{PhysRevC.91.014310,PhysRevLett.124.232501,c3st-tp13}. In such cases, reference-based many-body expansions may converge slowly because the omitted higher-rank excitations or induced higher-body operators become important; a quantitatively controlled description may then require very high truncation levels, such as IMSRG(4) or CCSDTQ, which are extremely computationally demanding. FCIQMC instead samples the wave function in the full configuration space of a chosen single-particle basis and symmetry sector, without imposing an excitation-rank truncation on the wave function. It can therefore treat closed- and open-shell nuclei in the same framework and provides a systematically improvable route to the FCI limit by increasing the walker population. These properties make FCIQMC attractive both as an accurate stochastic solver in large model spaces and as a benchmark for assessing the residual many-body uncertainties of truncated many-body expansion methods.

This paper is a companion to a Letter that reports the main physics results obtained with FCIQMC for finite nuclei~\cite{companion_prl}. Here we focus on the algorithmic and numerical details needed to support those discussions. We first present the nuclear FCIQMC formalism, including excitation generation, observable estimators, the initiator approximation, adaptive shift correction and RDM sampling. We then analyze the imaginary-time dynamics, benchmark the method against deterministic FCI in small model spaces, and quantify the residual finite-walker uncertainty in larger model spaces through walker-number convergence and controlled infinite-walker extrapolations. The calculation of radii from RDM-based pure estimators is discussed explicitly. Finally, an excited-state calculation of $^6$Li is shown as a representative demonstration of the multi-state FCIQMC extension.

\section{Method}
In this section we describe the FCIQMC formalism used for \textit{ab initio} nuclear structure calculations.

\subsection{Hamiltonian and many-body basis}
We begin with the intrinsic Hamiltonian for $A$ nucleons,
\begin{equation}
    \hat{H}=\frac{1}{A} \sum_{i<j}^A \frac{(\boldsymbol{p}_i-\boldsymbol{p}_j)^2}{2 m}
    +\sum_{i<j}^A \hat{V}^{ij}_{\mathrm{NN}} +\sum_{i<j<k}^A \hat{V}^{ijk}_{\mathrm{3N}},
\end{equation}
where the first term represents the intrinsic kinetic energy, whereas $\hat{V}_{\mathrm{NN}}$ and $\hat{V}_{\mathrm{3N}}$ denote nucleon-nucleon (NN) and three-nucleon (3N) interactions, respectively.

The single-particle basis is chosen as a spherical harmonic-oscillator (HO) basis, with a frequency $\hbar\omega$ and a quantum-number truncation $e=2n+l\leq e_\mathrm{max}$. The FCI basis consists of all SDs in the chosen symmetry sector, specified by particle number, parity $P$ and total angular-momentum projection $M$. No further configuration truncation, such as an $N_\mathrm{max}$ or excitation-rank truncation, is imposed in the FCIQMC calculations. In this basis, the wave function is expanded as
\begin{equation}
    \ket{\Psi} = \sum_{i} C_i \ket{D_i},
\end{equation}
where $C_i$ are the CI coefficients to be calculated. The Hamiltonian matrix elements
\begin{equation}
    H_{ij}=\braket{D_i}{\hat{H}|D_j}
\end{equation}
are evaluated on-the-fly during the stochastic propagation, so the many-body Hamiltonian matrix is never stored explicitly.

\subsection{Imaginary-time projection}
FCIQMC is a stochastic projector method based on the imaginary-time Schr\"odinger equation
\begin{equation}
-\dfrac{\mathrm{d}}{\mathrm{d}\tau} \ket{\Psi(\tau)} = (\hat{H}-E_0)\ket{\Psi(\tau)},
\end{equation}
where $E_0$ is the lowest eigenvalue of $\hat{H}$ in the targeted symmetry sector, and $\tau$ is the imaginary time. Equivalently, the imaginary-time projector is
\begin{equation}
    \hat{P}(\tau) = e^{-\tau(\hat{H}-E_0)}.
\end{equation}
For any initial state with nonzero overlap with the ground
state in the same symmetry sector, the ground state can be projected out in the long-time limit,
\begin{equation}
    \ket{\Psi_0} \propto \lim_{\tau\to\infty}\hat{P}(\tau)\ket{\Psi(0)}.
\end{equation}
In practice, FCIQMC applies the short-time propagator~\cite{Booth2009,Booth2010,Booth2011}
\begin{equation}
\label{equ:propagator}
    \ket{\Psi(\tau+\Delta\tau)} = \big[1 -\Delta\tau (\hat{H}-S)\big] \ket{\Psi(\tau)},
\end{equation}
where $S$ is a dynamically adjusted energy offset that replaces the unknown $E_0$. In FCIQMC, $S$ is usually referred to as the shift. The time step should be small enough for stable projection, satisfying $\Delta\tau<2/(E_\mathrm{max}-E_0)$, where $E_\mathrm{max}$ is the largest eigenvalue in the finite model space. Since the short-time propagator has the same eigenvectors as $\hat{H}$, different values of $\Delta\tau$ change the convergence rate but do not introduce a time-step bias.

\subsection{Stochastic algorithm}\label{subsec:algorithm}
The CI coefficients are sampled by signed walkers. Each walker carries a sign $s_a=\pm 1$ and occupies one determinant. The signed walker population on $\ket{D_i}$ is
\begin{equation}
    N_i(\tau) = \sum_{a\in\ket{D_i}} s_a,
\end{equation}
and the total walker population is
\begin{equation}
    N_\mathrm{w}(\tau) = \sum_{i} |N_i(\tau)|.
\end{equation}
The walker distribution represents the stochastic wave function,
\begin{equation}
    \ket{\Psi(\tau)} = \sum_{i} N_i(\tau) \ket{D_i},
\end{equation}
up to an overall normalization. Inserting this expansion into Eq.~(\ref{equ:propagator}) gives the discrete master equation
\begin{equation}
\label{equ:master}
    N_i(\tau+\Delta\tau) = \big[1-\Delta\tau(H_{ii}-S)\big]N_i(\tau) - \Delta\tau\sum_{j\neq i} H_{ij}N_j(\tau).
\end{equation}

The stochastic realization of Eq.~(\ref{equ:master}) consists of three steps:
\begin{enumerate}
\item \textit{Spawning}: For each walker on $\ket{D_i}$ with sign $s_i$, a connected determinant $\ket{D_f}$ is generated with a known probability $p_\mathrm{gen}(f|i)$. A child walker is then spawned with probability
\begin{equation}
    p_\mathrm{s}(f|i) = \dfrac{\Delta \tau |H_{fi}|}{p_\mathrm{gen}(f|i)},
\end{equation}
and sign
\begin{equation}
    s_f = -s_i\, \mathrm{sgn}(H_{fi}).
\end{equation}

\item \textit{Death/cloning}: For each walker on $\ket{D_i}$ with sign $s_i$, we first calculate its death probability
\begin{equation}
    p_\mathrm{death}(i) = \Delta\tau |H_{ii}-S|.
\end{equation}
If $H_{ii}>S$, the walker is removed with probability $p_\mathrm{death}(i)$; if $H_{ii}<S$, an additional walker with the same sign is cloned on $\ket{D_i}$.

\item \textit{Annihilation}: After spawning and death/cloning, all walkers on the same determinant are collected and walkers of opposite signs are canceled exactly. The remaining signed population is the net coefficient sample $N_i$. This step is essential for establishing a stable fermionic sign structure in the sampled wave function~\cite{Booth2009,Booth2010,Booth2011}.
\end{enumerate}

A typical FCIQMC calculation is divided into three stages. First, in the initialization stage, walkers are placed on one or several determinants with the desired quantum numbers. A single determinant with a low diagonal energy is sufficient in principle, while a multi-configurational initial state can reduce equilibration time for open-shell or strongly collective systems~\cite{Blunt2015_2}. Second, in the fixed-shift stage, $S$ is kept fixed, usually near the diagonal energy of the reference determinant, so that $N_\mathrm{w}$ grows until a target population $N_\mathrm{target}$ is reached. Finally, in the variable-shift stage, $S$ is updated every $N_S$ time steps according to
\begin{equation}
S(\tau) = S(\tau - N_S\Delta \tau) - \dfrac{\xi}{N_S\Delta \tau} \ln{\dfrac{N_\mathrm{w}(\tau)}{N_\mathrm{w}(\tau-N_S\Delta \tau)}},
\end{equation}
where $\xi$ is a damping parameter. This feedback keeps the walker population fluctuating around $N_\mathrm{target}$. Observables are accumulated only after the walker distribution and the shift have equilibrated. We adapt $N_S=10$ and $\xi=0.1$ in all calculations.

\subsection{Observable estimators}\label{subsec:obs}
Projected estimators exploit the fact that, for an exact eigenstate, the action of the Hamiltonian can be projected onto a reference state. For the energy, using a reference determinant $\ket{D_0}$ with nonzero overlap with the sampled state gives
\begin{equation}
E_\mathrm{proj} = \dfrac{\braket{D_0}{\hat{H}|\Psi(\tau)}}{\braket{D_0}{\Psi(\tau)}} = \sum_{i} H_{0i} \dfrac{N_i(\tau)}{N_0(\tau)}.
\end{equation}
The reference determinant is normally chosen as the most populated determinant. The projected estimator has a very small computational cost because it only requires the population of determinants directly connected to $\ket{D_0}$ by the Hamiltonian, i.e., up to double excitations for a two-body Hamiltonian. More generally, a projected estimator is valid for an operator $\hat{O}$ whose action
on the targeted non-degenerate eigenstate is proportional
to that eigenstate, in particular for operators commuting with $\hat{H}$ within the relevant symmetry sector such as the square of total angular-momentum operator $\hat{J}^2$.

The trial estimator~\cite{PhysRevLett.109.230201,Blunt2015_2,Guther2020} reduces statistical fluctuations by replacing the single reference determinant with a compact trial wave function,
\begin{equation}
O_\mathrm{trial}=\dfrac{\braket{\psi_T}{\hat{O}|\Psi(\tau)}}{\braket{\psi_T}{\Psi(\tau)}}.
\end{equation}
The trial state $\ket{\psi_T}$ is obtained by diagonalizing $\hat{H}$ in a trial space $\mathcal{T}$ spanned by the $N_T$ most populated determinants. This estimator is especially useful for the energy of multi-reference systems, where no single determinant dominates the wave function.

For observables that do not commute with the Hamiltonian, such as radius operators, projected or trial estimators are not equal to the desired expectation value. We therefore use the pure estimator~\cite{Booth2012_3,Overy2014,Blunt2017}
\begin{equation}
\label{equ:pure}
    O_{\mathrm{pure}}
    =\frac{\braket{\Psi(\tau)}{\hat{O}|\Psi(\tau)}}
    {\braket{\Psi(\tau)}{\Psi(\tau)}}.
\end{equation}
For a sampled wave function, the pure energy estimator is variational, while the point-proton radius and charge radius are obtained from pure expectation values of the corresponding radius operators.

After equilibration, time-series estimators such as the shift, projected estimator and trial estimator are accumulated, and their statistical uncertainties are estimated with blocking analysis~\cite{10.1063/1.457480,code-pyblock}, which accounts for auto-correlation in the imaginary-time series. The statistical uncertainties of RDM-based pure estimators are estimated from the residual RDMs discussed below.

\subsection{Excitation generation}
The spawning step requires an excitation generator that selects every connected determinant $\ket{D_f}$ from $\ket{D_i}$ with a nonzero and exactly computable probability $p_{\mathrm{gen}}(f|i)$. For a Hamiltonian up to two-body level, determinants are connected by single and double excitations, so the generator first chooses the excitation rank with normalized probabilities
\begin{equation}
    p_{\mathrm{single}}+p_{\mathrm{double}}=1.
\end{equation}
We take $p_{\mathrm{single}}=0.1$ and $p_{\mathrm{double}}=0.9$ in all calculations.

In the uniform single-excitation generator, an occupied orbital $p$ in $\ket{D_i}$ is selected with probability $1/A$. The target orbital $q$ is then selected uniformly from the unoccupied orbitals in the same allowed single-excitation channel, i.e., the set of orbitals for which replacing $p$ by $q$ preserves the conserved many-body quantum numbers and can give a nonzero Hamiltonian matrix element. If the number of such orbitals is $N_i(q|p)$, the generation probability is
\begin{equation}
    p_{\mathrm{gen}}(f|i)
    =p_{\mathrm{single}}\binom{A}{1}^{-1}\frac{1}{N_i(q|p)}.
\end{equation}
The single excitation gives the final determinant as
\begin{equation}
    \ket{D_i}\to \hat{a}_q^\dagger \hat{a}_p \ket{D_i} \to \ket{D_f}.
\end{equation}

For a uniform double excitation, an occupied pair $(p,q)$ is selected with probability $1/\binom{A}{2}$. The target pair $(r,s)$ is selected uniformly from the unoccupied two-body states in the same two-body channel, defined by the conserved angular-momentum projection, parity and isospin projection of the pair. If this number is $N_i(rs|pq)$, then
\begin{equation}
    p_{\mathrm{gen}}(f|i)
    =p_{\mathrm{double}}\binom{A}{2}^{-1}
    \frac{1}{N_i(rs|pq)}.
\end{equation}
The double excitation gives the final determinant as
\begin{equation}
    \ket{D_i}\to \hat{a}_r^\dagger \hat{a}_s^\dagger \hat{a}_p \hat{a}_q \ket{D_i} \to \ket{D_f}.
\end{equation}

Uniform generation is simple but inefficient for realistic nuclear Hamiltonians, whose off-diagonal matrix elements vary over a wide range. It spends many attempts on weak or vanishing couplings, leading to small spawning probabilities and high rejection rates. We therefore use a pre-computed heat-bath generator~\cite{Walker1977,Holmes2016,Holmes2016_2,Guther2020} when memory permits. For a double excitation, after selecting the occupied pair $(p,q)$, the target pair $(r,s)$ is drawn with probability
\begin{equation}
    p(rs|pq)=
    \frac{|H_{pq}^{rs}|}
    {\sum_{r's'} |H_{pq}^{r's'}|},
\end{equation}
where $H_{pq}^{rs}$ is the antisymmetrized two-body matrix element and the sum is taken over target pairs in the corresponding two-body channel. This choice strongly improves the quality of the generated excitations by preferentially sampling large-matrix-element contributions. Because the probability distributions are pre-computed and can be used directly, it also accelerates the excitation-generation step itself. The resulting lower rejection rate and faster generation give a substantial efficiency gain compared with uniform excitation generation, at the cost of storing the heat-bath tables~\cite{Walker1977}.

\subsection{Initiator approximation and adaptive shift}
\label{subsec:initiator}
In the original FCIQMC algorithm, a stable sign structure appears only above a system-dependent critical walker population~\cite{Booth2009,Booth2010,Booth2011,PhysRevB.90.155130}. For large nuclear configuration spaces, reaching this population with the unbiased algorithm can be impractical. This critical walker population is a manifestation of the sign problem of the FCIQMC method, and height of the plateau an indication of the severity of the sign problem. The initiator approximation (i-FCIQMC)~\cite{Cleland2010,Cleland2011} was introduced to overcome this bottleneck by restricting spawning from weakly occupied determinants, and practically removes the sign problem. The initiator approximation has become one of the most successful and widely used improvements of FCIQMC, enabling applications to systems far beyond the reach of the original algorithm. A determinant $\ket{D_i}$ is called an initiator if
\begin{equation}
    |N_i|> n_{\mathrm{init}},
\end{equation}
where $n_{\mathrm{init}}$ is the initiator threshold. Spawns from initiators are always allowed. Spawns from non-initiators are allowed only if the target determinant is already occupied; otherwise they are rejected.

Schematically, the initiator rule replaces the off-diagonal Hamiltonian by a stochastic, population-dependent matrix
\begin{equation}
\tilde{H}_{ij}(\tau)=
\begin{cases}
0, & |N_j(\tau)|\leq n_{\mathrm{init}} \ \mathrm{and}\ N_i(\tau)=0,\\
H_{ij}, & \mathrm{otherwise},
\end{cases}
\end{equation}
where $\ket{D_j}$ is the spawning determinant and $\ket{D_i}$ is the target determinant. The initiator space is updated dynamically during the simulation. This approximation practically removes the sign problem at the expense of a bias. The bias, however, can be reduced to essentially zero (i.e. achieving FCI quality results) by increasing the walker population:
\begin{equation}
\lim_{N_{\mathrm{w}}\to\infty}\mathrm{i}\text{-}\mathrm{FCIQMC} =\mathrm{FCIQMC}=\mathrm{FCI},
\end{equation}
as all determinants become initiators in this limit.

In practice, for many systems, the FCI limit is reached with the initiator method with a much smaller walker population than with the full FCIQMC method. The initiator method has been further refined with the adaptive shift (AS) method~\cite{Ghanem2019,Ghanem2020} by assigning a local shift $S_i(\tau)$ to non-initiator determinants,
\begin{equation}
    S_i(\tau)=\Delta+f_i\left[S(\tau)-\Delta\right],
\end{equation}
while initiators have $f_i=1$ and therefore $S_i=S$. The factor $f_i$ measures how much the spawning of determinant $\ket{D_i}$ is accepted by the initiator criterion:
\begin{equation}
    f_i=\frac{\sum_{\mathrm{accepted}} w_{ij}}
    {\sum_{\mathrm{all}} w_{ij}},
\end{equation}
where the sums run over attempted spawns from $\ket{D_i}$ and
\begin{equation}
    w_{ij}=\frac{|H_{ij}|}{H_{jj}-E_{\mathrm{ref}}}
\end{equation}
is a weight from perturbation theory. In actual calculations the unknown $E_{\mathrm{ref}}\simeq E_0$ is replaced by the current energy estimate, typically the shift or projected energy. We parameterize the offset parameter as $\Delta=\tilde{\Delta}S(\tau)$. The offset controls the strength of the adaptive-shift correction; $\tilde{\Delta}=0$ gives the full strength, while larger offsets reduce the correction.

In the large-walker limit, all determinants become initiators, so $f_i\to 1$ and
\begin{equation}
    \lim_{N_{\mathrm{w}}\to\infty}\mathrm{AS}\text{-}\mathrm{FCIQMC}
    =\mathrm{i}\text{-}\mathrm{FCIQMC}.
\end{equation}
Thus the AS method accelerates the walker-population convergence but preserves the exact FCI limit. The total walker number $N_\mathrm{w}$ remains the primary parameter that systematically controls the residual initiator error. In addition, comparing calculations with different offset parameters $\tilde{\Delta}$ and initiator thresholds $n_{\mathrm{init}}$ provides an important practical check that the remaining systematic bias has been removed within the quoted uncertainty. Moreover, the replica-sampling for pure estimators (discussed in next subsection) provides another independent way to compute energies, and the totality of these different methods allow us to assess the accuracy and expected errors in energies and other properties within FCIQMC method.

\subsection{Reduced-density-matrix sampling}
\label{subsec:rdm}
Pure estimators are evaluated through reduced density matrices (RDMs). This is essential in large FCIQMC calculations: evaluating Eq.~(\ref{equ:pure}) directly from the stochastic wave function would require an explicit double sum over a prohibitively large number of determinant pairs. RDM sampling converts this problem into an accumulation of density-matrix elements during the same spawning events already generated in the FCIQMC propagation~\cite{Overy2014,Blunt2017}. For an operator containing up to two-body terms,
\begin{equation}
    \hat{O}=O^{(0)}
    +\sum_{pq}O^{(1)}_{pq}a_p^\dagger a_q
    +\frac{1}{4}\sum_{pqrs}O^{(2)}_{pqrs}
    a_p^\dagger a_q^\dagger a_s a_r,
\end{equation}
where $O^{(2)}_{pqrs}$ are antisymmetrized two-body matrix elements, the pure expectation value is
\begin{equation}
    O_{\mathrm{pure}}=O^{(0)}
    +\sum_{pq}O^{(1)}_{pq}\gamma_{pq}
    +\sum_{p<q,r<s}O^{(2)}_{pqrs}\Gamma_{pqrs},
\end{equation}
with one- and two-body RDMs
\begin{equation}
     \gamma_{pq}=\braket{\Psi}{a_p^\dagger a_q|\Psi},
\end{equation}
\begin{equation}
    \Gamma_{pqrs}=\braket{\Psi}{a_p^\dagger a_q^\dagger a_s a_r|\Psi}
\end{equation}
up to a normalization factor. The 1-RDM can be obtained from the 2-RDM through
\begin{equation}
    \gamma_{pq}=\frac{1}{A-1}\sum_a \Gamma_{paqa},
\end{equation}
so the 2-RDM is sufficient for all one- and two-body observables.

Using the wave function in the FCI basis, $\ket{\Psi}=\sum_i C_i\ket{D_i}$, the 2-RDM is
\begin{equation}
    \Gamma_{pqrs}= \sum_{ij} C_i^* C_j \braket{D_i}{a_p^\dagger a_q^\dagger a_s a_r|D_j}.
\end{equation}
For $p<q$, the diagonal element of the 2-RDM is
\begin{equation}
    \Gamma_{pqpq}= \sum_{p,q\in \ket{D_i}} |C_i|^2.
\end{equation}
Off-diagonal elements of 2-RDM are nonzero only when $\ket{D_i}$ and $\ket{D_j}$ differ by the corresponding one- or two-body excitation, in which case the operator matrix element gives a fermionic phase $\pm 1$.

Directly replacing $C_i$ by the instantaneous walker population $N_i(\tau)$ gives a biased RDM because the RDM is quadratic in stochastic amplitudes~\cite{Overy2012,Overy2014,Blunt2017}. To remove this bias we use two independent replicas, with walker populations $N_i^{(1)}$ and $N_i^{(2)}$, propagated with independent random numbers. Diagonal elements are accumulated in an accumulator $\widetilde{\Gamma}$ from products of independent populations,
\begin{equation}
    \Delta\widetilde{\Gamma}_{pqpq}=
    N_i^{(1)}(\tau)N_i^{(2)}(\tau),
    \quad p<q \text{ and } p,q\in \ket{D_i}.
\end{equation}
For an off-diagonal connection sampled in replica 1 from $\ket{D_j}$ to $\ket{D_i}$, we accumulate the contribution
\begin{equation}
    \Delta\widetilde{\Gamma}_{pqrs}=
    \dfrac{N_i^{(2)}(\tau)\,\mathrm{sgn}\big(N_j^{(1)}(\tau)\big)
    \braket{D_i}{a_p^\dagger a_q^\dagger a_s a_r|D_j}}{p_{\mathrm{gen}}(i|j)}.
\end{equation}

After sampling, the unnormalized accumulator $\widetilde{\Gamma}$ is converted to the normalized $\Gamma$ by enforcing
\begin{equation}
    \sum_{p<q}\Gamma_{pqpq}=\frac{A(A-1)}{2}.
\end{equation}
Stochastic sampling does not produce an exactly Hermitian RDM, so we symmetrize it by
\begin{equation}
    \bar{\Gamma}_{pqrs}
    =\frac{\Gamma_{pqrs}+\Gamma_{rspq}^*}{2}.
\end{equation}
The antisymmetric residual 2-RDM,
\begin{equation}
    \delta\Gamma_{pqrs}
    =\frac{\Gamma_{pqrs}-\Gamma_{rspq}^*}{2},
\end{equation}
provides a useful diagnostic of the statistical noise and vanishes in the infinite-sampling limit. In this work, the statistical uncertainty of each RDM-based pure estimator is estimated from the corresponding contraction with this residual RDM. The RDM sampling therefore gives access to pure estimators for energies, radii and other one- or two-body observables with almost no additional computational cost, since the determinant connections, matrix elements and generation probabilities are already available in the spawning algorithm. This is one of the practical advantages of FCIQMC for computing observables beyond the projected energy.

\subsection{Excited states}
\label{subsec:ex}
Although the calculations emphasized in this work focus on ground states, FCIQMC can also target excited states~\cite{Blunt2015_2}. States in different symmetry sectors are obtained by separate calculations. To obtain several states in the same symmetry sector, one propagates $K$ walker ensembles in parallel and orthogonalizes them after each iteration. For state index $\nu$, the Gram-Schmidt step is
\begin{equation}
    \ket{\Psi_\nu}\leftarrow
    \ket{\Psi_\nu}
    -\sum_{\mu<\nu}
    \frac{\braket{\Psi_\mu}{\Psi_\nu}}
    {\braket{\Psi_\mu}{\Psi_\mu}}
    \ket{\Psi_\mu}.
\end{equation}
In the long-time limit, this procedure samples the $K$ lowest eigenstates in the chosen symmetry sector.

\section{Results and discussion}
The optimized chiral NN interaction N$^2$LO$_\mathrm{opt}$~\cite{PhysRevLett.110.192502} is used throughout this work, with the HO frequency $\hbar\omega=20$ MeV. Unless otherwise stated, the calculations use AS-FCIQMC with $N_S=10$, $\xi=0.1$, $n_\mathrm{init}=3$ and $\tilde{\Delta}=0$.

For comparison, we also show results from several other \textit{ab initio} many-body methods. NCSM diagonalizes the many-body Hamiltonian matrix in an $N_\mathrm{max}$-truncated many-body space deterministically~\cite{Barrett2013}. MBPT(2) and MBPT(3) denote many-body perturbation theory truncated at second and third order~\cite{10.3389/fphy.2020.00164}, respectively. CCSD retains single and double cluster amplitudes, while CCSDT-3~\cite{Noga1987} adds an approximate treatment of triple excitations. IMSRG(2) denotes the in-medium similarity renormalization group truncated at the normal-ordered two-body level, and IMSRG(3f$_2$)+T~\cite{PhysRevC.110.044317} denotes an approximate IMSRG(3) scheme with perturbative triples correction. The MBPT and CC results are reported only for closed-shell nuclei, whereas IMSRG results are obtained with the single-reference or valence-space formulation~\cite{PhysRevLett.118.032502,PhysRevLett.106.222502} as appropriate.

\subsection{Imaginary-time dynamics}
\begin{figure}[t]
\centering
\resizebox{0.48\textwidth}{!}{
\includegraphics{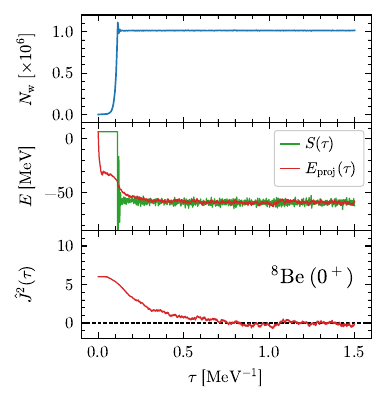}}
\caption{Representative FCIQMC imaginary-time evolution for the $^8$Be $J^\pi=0^+$ ground state in the $e_\mathrm{max}=10$ model space with $N_\mathrm{w}=10^6$. From top to bottom, the panels show the walker population $N_\mathrm{w}(\tau)$, the shift $S(\tau)$ and projected energy $E_\mathrm{proj}(\tau)$, and the projected estimator of $\hat{J}^2$.}
\label{fig:evolution_example}
\end{figure}

\begin{figure}[t]
\centering
\resizebox{0.48\textwidth}{!}{
\includegraphics{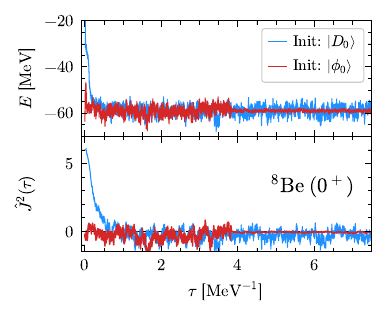}}
\caption{Effect of multi-configurational initialization and trial estimators for the same $^8$Be calculation. The state $\ket{\phi_0}$ is obtained from a preliminary $N_\mathrm{w}=10^3$ calculation and is compared with single-determinant initialization from $\ket{D_0}$. After $\tau\simeq 3.8$ MeV$^{-1}$, the $\ket{\phi_0}$ calculation switches to trial estimators with $N_T=100$.}
\label{fig:evolution_comparison}
\end{figure}

Figure~\ref{fig:evolution_example} shows a representative AS-FCIQMC calculation for the $J^\pi=0^+$ ground state of $^8$Be in the $e_\mathrm{max}=10$ model space. We monitor the total walker population $N_\mathrm{w}$, the shift $S$, the projected energy estimator $E_\mathrm{proj}$ and the projected estimator of the square of the total angular momentum,
\begin{equation}
    \hat{J}^2 = \bigg(\sum_{i=1}^A \hat{j}_i\bigg)^2,
\end{equation}
where $\hat{j}_i$ is the total angular-momentum operator of the $i$-th nucleon. For a state with good total angular momentum, $\langle\hat{J}^2\rangle=J(J+1)$; therefore the target $0^+$ state is expected to give $\langle\hat{J}^2\rangle\simeq 0$ after equilibration.

The calculation starts from a single positive walker on the reference determinant $\ket{D_0}$. During the initial fixed-shift stage, $0<\tau\lesssim 0.1$ MeV$^{-1}$, the walker population grows exponentially until it reaches the target value $N_\mathrm{w}=10^6$. The shift is then updated dynamically to stabilize the population around this target. In the following equilibration stage, $0.1$ MeV$^{-1}\lesssim\tau\lesssim 0.8$ MeV$^{-1}$, the sampled wave function is progressively projected toward the ground state. After equilibration, $S(\tau)$ and $E_\mathrm{proj}(\tau)$ fluctuate around a common energy plateau, while $\langle\hat{J}^2\rangle$ fluctuates around zero, and then the time series is used for statistical analysis.

Two choices are especially useful for improving the efficiency of such calculations: a multi-configurational initial state and/or a trial estimator. To illustrate their effect, we first perform a small preliminary calculation with $N_\mathrm{w}=10^3$ and store the final walker distribution as $\ket{\phi_0}$. Figure~\ref{fig:evolution_comparison} compares the baseline calculation initialized from $\ket{D_0}$ with a calculation initialized from $\ket{\phi_0}$. The multi-configurational initialization reduces the equilibration time for both the projected energy and $\langle\hat{J}^2\rangle$. In the $\ket{\phi_0}$-calculation, after $\tau\simeq 3.8$ MeV$^{-1}$, the energy and $\hat{J}^2$ estimators are switched from projected estimators to trial estimators with a trial-space size $N_T=100$, which further reduces statistical fluctuations.

In the calculations below, we therefore always use the multi-configurational initialization.

\subsection{Benchmark calculations}
\begin{figure}[t]
\centering
\resizebox{0.48\textwidth}{!}{
\includegraphics{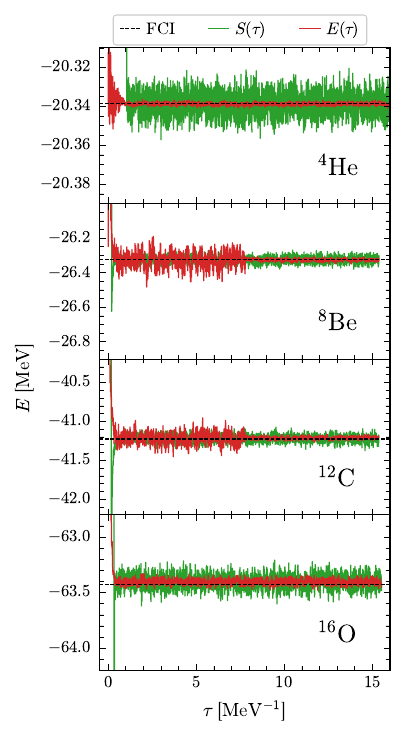}}
\caption{Imaginary-time evolution of the ground-state energy estimates and shifts for $^4$He, $^8$Be, $^{12}$C and $^{16}$O in the $e_\mathrm{max}=2$ model space using $N_\mathrm{w}=10^7$. The horizontal dashed lines denote the FCI energies from direct diagonalization.}
\label{fig:evolution_emax2}
\end{figure}

We first benchmark the FCIQMC calculations in the $e_\mathrm{max}=2$ model space for the ground states of $^4$He, $^8$Be, $^{12}$C and $^{16}$O. This space is small enough that exact FCI results can be obtained by direct diagonalization, while still allowing us to test the stochastic propagation, estimator choices and RDM sampling used in the larger calculations below. Figure~\ref{fig:evolution_emax2} shows the imaginary-time evolution of the shift and energy estimate for each nucleus with $N_\mathrm{w}=10^7$. For $^4$He and $^{16}$O, $E(\tau)$ is the projected estimator throughout the calculation. For $^8$Be and $^{12}$C, the estimator is switched from the projected estimator to a trial estimator with $N_T=100$ at $\tau\simeq 8$ MeV$^{-1}$. After equilibration, the FCIQMC energy estimates fluctuate around the corresponding FCI values, shown by the horizontal dashed lines.

\begin{table*}
    \caption{Ground-state energies ($E$) and squared point-proton radii ($R_\mathrm{p}^2$) of $^4$He, $^8$Be, $^{12}$C and $^{16}$O in the $e_\mathrm{max}=2$ model space. The FCIQMC values are pure estimators from sampled 2-RDMs using $N_\mathrm{w}=10^7$; parentheses denote statistical uncertainties.}
    \label{tab:emax2}
    \begin{ruledtabular}
    \begin{tabular}{ p{2.3cm} p{1.6cm} p{1.6cm} p{1.6cm} p{1.6cm} p{1.6cm} p{1.6cm} p{1.6cm} p{1.6cm} }
    \multirow{2}{*}{Method}
    & \multicolumn{2}{c}{$^4$He}
    & \multicolumn{2}{c}{$^8$Be}
    & \multicolumn{2}{c}{$^{12}$C}
    & \multicolumn{2}{c}{$^{16}$O} \\
    \cline{2-3} \cline{4-5} \cline{6-7} \cline{8-9}
    &
    \rule{0pt}{2.5ex}$E$ [MeV] & $R_\mathrm{p}^2$ [fm$^2$]
    & $E$ [MeV] & $R_\mathrm{p}^2$ [fm$^2$]
    & $E$ [MeV] & $R_\mathrm{p}^2$ [fm$^2$]
    & $E$ [MeV] & $R_\mathrm{p}^2$ [fm$^2$] \\[2pt]
    \noalign{\vskip 1pt}
    \hline
    \hline
    \noalign{\vskip 1pt}
    \rule{0pt}{2.5ex}FCI       & $-20.34$ & $2.16$ & $-26.33$ & $4.21$ & $-41.23$ & $4.48$ & $-63.42$ & $4.72$  \\
    FCIQMC    & $-20.34(0)$ & $2.16(0)$ & $-26.33(0)$ & $4.21(0)$ & $-41.21(1)$ & $4.48(0)$ & $-63.40(4)$ & $4.72(0)$ \\
    MBPT(2)     & $-19.01$ & $1.93$ &  &  & $-68.94$ & $4.11$ & $-61.53$ & $4.56$ \\
    MBPT(3)     & $-19.40$ & $2.04$ &  &  & $-39.31$ & $4.20$ & $-62.65$ & $4.65$ \\
    CCSD      & $-19.74$ & $2.14$ &  &  & $-34.74$ & $4.32$ & $-62.59$ & $4.73$ \\
    CCSDT-3   & $-19.93$ & $2.17$ &  &  & $-40.90$ & $4.39$ & $-63.29$ & $4.73$ \\
    IMSRG(2)    & $-19.79$ & $2.09$ & $-26.34$ & $4.19$ & $-41.73$ & $4.43$ & $-63.73$ & $4.76$ \\
    IMSRG(3f$_2$)+T  & $-19.85$ & $2.22$ & $-24.25$ & $4.16$ & $-40.43$ & $4.44$ & $-63.24$ & $4.72$
    \end{tabular}
    \end{ruledtabular}
\end{table*}

We then use the same benchmark calculations to test RDM-based pure estimators, which are required for non-commuting observables such as radii. During the last 7.5 MeV$^{-1}$ of imaginary-time evolution, the 2-RDM is sampled using the procedure described in Sec.~\ref{subsec:rdm}. The sampled 2-RDM gives pure estimators for the energy, $E_\mathrm{pure}$, and for the expectation value of the squared point-proton radius operator, $R_\mathrm{p}^2$, with statistical uncertainties estimated from the antisymmetric residual 2-RDMs $\delta\Gamma$. The results are summarized in Table~\ref{tab:emax2} and compared with FCI, together with results from other many-body expansion methods, including MBPT(2), MBPT(3), CCSD, CCSDT-3, IMSRG(2) and IMSRG(3f$_2$)+T. The agreement between FCIQMC and FCI for both energies and radii shows that, in this benchmark space, the walker distribution and the sampled RDM reproduce the exact expectation values within statistical uncertainties.

\begin{figure}[t]
\centering
\resizebox{0.48\textwidth}{!}{
\includegraphics{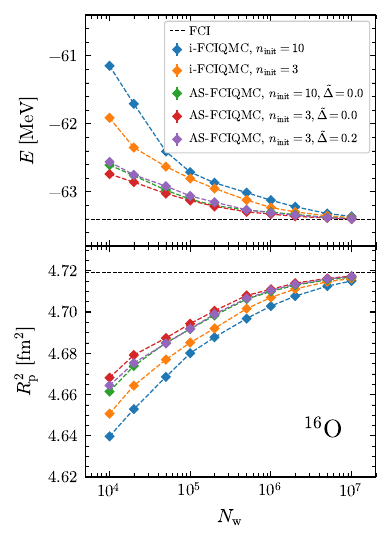}}
\caption{Walker-number convergence of the $^{16}$O ground-state energy and squared point-proton radius in the $e_\mathrm{max}=2$ model space for different initiator prescriptions. The horizontal dashed lines denote the FCI results.}
\label{fig:bias_emax2}
\end{figure}

Finally, we examine the convergence of the initiator bias in this small space, using $^{16}$O as a representative case. We repeat the calculations with several setups: standard i-FCIQMC with $n_\mathrm{init}=3$ and 10; AS-FCIQMC with $n_\mathrm{init}=3$ and 10 at $\tilde{\Delta}=0$; and AS-FCIQMC with $n_\mathrm{init}=3$ at $\tilde{\Delta}=0.2$. As discussed in Sec.~\ref{subsec:initiator}, these choices must approach the same FCI limit as $N_\mathrm{w}\to\infty$, although their finite-$N_\mathrm{w}$ biases can differ. Figure~\ref{fig:bias_emax2} shows that the energy and radius systematically converge toward the FCI results with increasing walker number. In these calculations the radius approaches the FCI value from below. The adaptive-shift correction substantially reduces the finite-walker bias and weakens the dependence on the initiator threshold. Therefore, comparisons among different $n_\mathrm{init}$ and $\tilde{\Delta}$ choices provide a useful diagnostic for residual initiator bias, which will be used in the larger model spaces where direct FCI benchmarks are unavailable.

\subsection{Convergence in the large model space}

\begin{figure*}[t]
    \centering
    \includegraphics[width=0.8\textwidth]{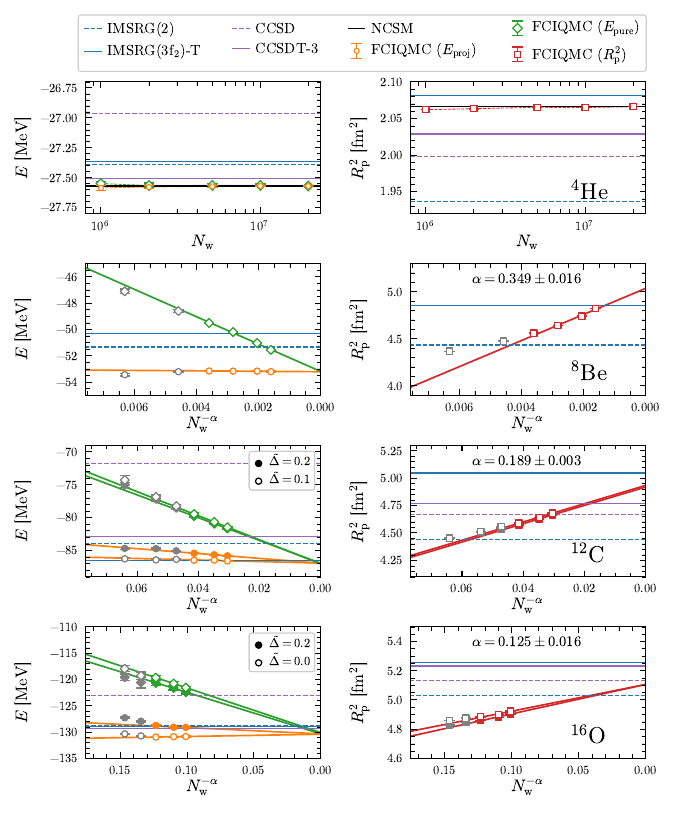}
    \caption{Walker-number convergence and infinite-walker extrapolation of the ground-state energy and squared point-proton radius in the $e_\mathrm{max}=10$ model space. The top row shows the direct convergence of $^4$He as functions of $N_\mathrm{w}$, using $N_\mathrm{w}=10^6$--$2\times10^7$. The lower rows show the extrapolations for $^8$Be, $^{12}$C and $^{16}$O as functions of $N_\mathrm{w}^{-\alpha}$, with $\alpha$ determined by requiring consistency between the extrapolated projected and pure energy estimators. The walker numbers are $2\times10^6$--$10^8$ for $^8$Be and $^{12}$C, and $5\times10^6$--$10^8$ for $^{16}$O. Gray markers denote walker populations not included in the linear fits. Horizontal lines show NCSM, when available, and selected many-body expansion results.}
    \label{fig:bias_emax10}
\end{figure*}

We now present calculations in the large $e_\mathrm{max}=10$ model space using AS-FCIQMC with $n_\mathrm{init}=3$. We first examine $^4$He, for which the calculation gets converged directly with increasing walker number. As shown in the top row of Fig.~\ref{fig:bias_emax10}, the projected energy and the RDM-based pure energy approach the same value from opposite sides, and both the energy and $R_\mathrm{p}^2$ agree with the available NCSM result once $N_\mathrm{w}$ reaches a few million. This case provides a direct validation of the large-space FCIQMC calculation and of the RDM sampling procedure.

The heavier nuclei, $^8$Be, $^{12}$C and $^{16}$O, show a different but systematic pattern. For these nuclei, available NCSM results at $\hbar\omega=20$ MeV show slow convergence for energies and radii, making reliable $N_\mathrm{max}$-extrapolated results impossible. As shown in the bottom three rows of Fig.~\ref{fig:bias_emax10}, for FCIQMC, the projected energy estimator converges rapidly with $N_\mathrm{w}$ and is already stable within small statistical fluctuations at the largest walker populations. In contrast, the RDM-based pure energy and $R_\mathrm{p}^2$ retain a visible residual walker-number dependence. This slower convergence is understandable because pure estimators probe the sampled wave function more globally through the RDM, whereas the projected estimator requires only the amplitudes of determinants connected by the Hamiltonian to the reference determinant. Consequently, the energy can be determined with a small residual many-body uncertainty from the rapidly-converging projected estimator, whereas the dominant residual uncertainty in these large-space calculations comes from the RDM-based radius.

This separation suggests a controlled way to estimate the remaining initiator bias in the RDM quantities. We adopt the walker-number extrapolation strategy developed in quantum-chemistry applications~\cite{Hosseini2024}, where the initiator bias is expected to follow a power-law asymptotic form in the large-$N_\mathrm{w}$ regime. We therefore fit the residual walker-number dependence of an observable $O$ as
\begin{equation}
    O(N_\mathrm{w}) = O_\infty + \lambda N_\mathrm{w}^{-\alpha}.
\end{equation}
We determine $\alpha$ by requiring the extrapolated pure energy $E_\mathrm{pure}$ to agree with the independently fast-converging projected energy $E_\mathrm{proj}$ within the statistical uncertainty. This energy-consistency condition determines both the central value and the allowed interval of $\alpha$. The same interval is then propagated to the extrapolation of $R_\mathrm{p}^2$, which gives the quoted uncertainty of the radius and, after conversion, of the charge radius.

The lower rows of Fig.~\ref{fig:bias_emax10} show this extrapolation for $^8$Be, $^{12}$C and $^{16}$O. For all three nuclei, the projected energy is nearly flat over the largest walker populations, while the pure energy approaches the same limit more slowly. For $^8$Be, requiring the extrapolated projected and pure energy estimators to be consistent gives $\alpha=0.349\pm0.016$. The radius is then extrapolated with this energy-constrained value of $\alpha$. The same procedure is applied to $^{12}$C and $^{16}$O. To test the robustness of the extrapolation, we perform two independent AS-FCIQMC calculations with different offset parameters: $\tilde{\Delta}=0.2$ and 0.1 for $^{12}$C; $\tilde{\Delta}=0.2$ and 0 for $^{16}$O. The extrapolated energies and radii obtained from these data sets agree within the estimated uncertainties. This $\tilde{\Delta}$-independence gives an additional check that the final results are controlled by the infinite-walker limit rather than by a particular choice of adaptive-shift parameter.

The final large-space results are summarized in Table~\ref{tab:emax10} and compared with other many-body methods. For $^4$He, the FCIQMC values are taken from the directly converged largest-walker calculation. For $^8$Be, $^{12}$C and $^{16}$O, the quoted central values and uncertainties are obtained from the $N_\mathrm{w}^{-\alpha}$ extrapolation described above. The comparison shows that FCIQMC captures correlations beyond the reach of the truncated many-body expansion methods in this model space.

The charge radii $R_\mathrm{ch}$ can be calculated from the squared point-proton radii using the formula~\cite{PhysRevC.102.051303,Hagen2015}:
\begin{equation}
R_\mathrm{ch} = \sqrt{R_\mathrm{p}^2 + r_{\mathrm{p}}^2 + \frac{N}{Z} r_\mathrm{n}^2 + \frac{3\hbar^2}{4 m_\mathrm{p}^2 c^2} + R_\mathrm{so}^2},
\end{equation}
where $r_{\mathrm{p}}=0.8409$ fm is the proton charge radius, $r_{\mathrm{n}}^2=-0.1155$ fm$^2$ is the neutron charge radius squared, the Darwin-Foldy term is $\frac{3\hbar^2}{4 m_\mathrm{p}^2 c^2}=0.033$ fm$^2$, and $R_\mathrm{so}^2$ is the spin-orbit correction~\cite{Mller2025,10.1093/ptep/ptac097}.

\begin{table*}
    \caption{Ground-state energies ($E$) and squared point-proton radii ($R_\mathrm{p}^2$) of $^4$He, $^8$Be, $^{12}$C and $^{16}$O in the $e_\mathrm{max}=10$ model space. The $^4$He FCIQMC values are directly converged, while the $^8$Be, $^{12}$C and $^{16}$O values are extrapolated to $N_\mathrm{w}\to\infty$; parentheses denote statistical uncertainties for $^4$He and estimated infinite-walker extrapolation uncertainties for the heavier nuclei.}
    \label{tab:emax10}
    \begin{ruledtabular}
    \begin{tabular}{ p{2.3cm} p{1.6cm} p{1.6cm} p{1.6cm} p{1.6cm} p{1.6cm} p{1.6cm} p{1.6cm} p{1.6cm} }
    \multirow{2}{*}{Method}
    & \multicolumn{2}{c}{$^4$He}
    & \multicolumn{2}{c}{$^8$Be}
    & \multicolumn{2}{c}{$^{12}$C}
    & \multicolumn{2}{c}{$^{16}$O} \\
    \cline{2-3} \cline{4-5} \cline{6-7} \cline{8-9}
    &
    \rule{0pt}{2.5ex}$E$ [MeV] & $R_\mathrm{p}^2$ [fm$^2$]
    & $E$ [MeV] & $R_\mathrm{p}^2$ [fm$^2$]
    & $E$ [MeV] & $R_\mathrm{p}^2$ [fm$^2$]
    & $E$ [MeV] & $R_\mathrm{p}^2$ [fm$^2$] \\[2pt]
    \noalign{\vskip 1pt}
    \hline
    \hline
    \noalign{\vskip 1pt}
    \rule{0pt}{2.5ex}NCSM       & $-27.57$ & $2.07$ &  &  &  &  &  &  \\
    FCIQMC    & $-27.57(00)$ & $2.07(00)$ & $-53.21(23)$ & $5.04(04)$ & $-86.92(34)$ & $4.92(03)$ & $-130.40(57)$ & $5.11(04)$ \\
    MBPT(2)     & $-28.27$ & $1.92$ &  &  & $-100.64$ & $4.95$ & $-132.37$ & $5.16$ \\
    MBPT(3)     & $-26.84$ & $1.93$ &  &  & $-93.53$ & $4.00$ & $-129.46$ & $5.00$ \\
    CCSD      & $-26.96$ & $2.00$ &  &  & $-71.73$ & $4.67$ & $-123.05$ & $5.13$ \\
    CCSDT-3   & $-27.51$ & $2.03$ &  &  & $-82.83$ & $4.77$ & $-129.33$ & $5.23$ \\
    IMSRG(2)    & $-27.39$ & $1.94$ & $-51.34$ & $4.43$ & $-83.90$ & $4.44$ & $-128.84$ & $5.03$ \\
    IMSRG(3f$_2$)+T  & $-27.37$ & $2.08$ & $-50.30$ & $4.86$ & $-86.46$ & $5.05$ & $-128.97$ & $5.25$
    \end{tabular}
    \end{ruledtabular}
\end{table*}

\begin{figure}[t]
    \centering
    \includegraphics[width=0.48\textwidth]{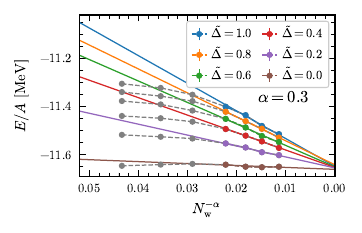}
    \caption{Walker-number extrapolation of the projected energy per particle in symmetric nuclear matter at $\rho=0.16$ fm$^{-3}$ with the $\Delta$N$^2$LO(450) interaction. The model space contains $A=76$ nucleons and 132 single-particle states. The walker numbers range $5\times10^4\,$--$\,5\times10^6$.}
    \label{fig:bias_matter}
\end{figure}

A similar large-walker power-law behavior is also observed in independent FCIQMC calculations of symmetric nuclear matter~\cite{hu2026fciqmc}. We consider matter at $\rho=0.16$ fm$^{-3}$ density on a periodic momentum-space lattice with 76 nucleons and 132 single-particle states, using the $\Delta$N$^2$LO(450) interaction~\cite{PhysRevC.102.054301}. As shown in Fig.~\ref{fig:bias_matter}, the projected energy per particle is approximately linear in $N_\mathrm{w}^{-\alpha}$ with $\alpha=0.3$ over the largest walker populations. Different choices of the adaptive-shift offset extrapolate to the same energy, while $\tilde{\Delta}\simeq 0$ gives the fastest convergence in this case. This independent many-body system therefore supports the use of the $N_\mathrm{w}^{-\alpha}$ form as a practical description of the residual initiator bias.

\subsection{Calculation of \texorpdfstring{$^6\text{Li}$}{} excited states}

Although the main focus of this work is on ground-state properties, the orthogonalized multi-state FCIQMC algorithm also provides a direct route to low-lying excited states, as described in Sec.~\ref{subsec:ex}. We illustrate this capability with the five lowest states of $^6$Li in the $e_\mathrm{max}=6$ model space. To suppress spurious excitations associated with center-of-mass motion, we add the Lawson term $\beta\hat{H}_\mathrm{c.m.}$ to the many-body Hamiltonian~\cite{PhysRevC.80.021306,Gloeckner1974,PhysRevC.82.034330,PhysRevC.102.034320} and use $\beta=4$ in both the FCIQMC and NCSM calculations.

\begin{figure}[t]
    \centering
    \includegraphics[width=0.48\textwidth]{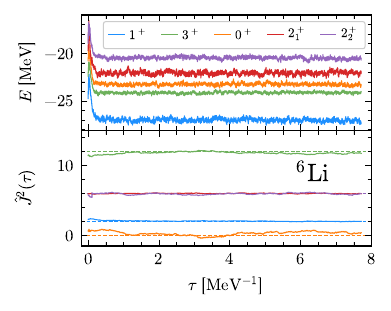}
    \caption{Imaginary-time evolution of the five lowest states of $^6$Li in the $e_\mathrm{max}=6$ model space using $N_\mathrm{w}=10^6$. The upper panel shows projected energy estimates, and the lower panel shows projected estimates of $\hat{J}^2$; dashed horizontal lines mark the corresponding $J(J+1)$ values.}
    \label{fig:evolution_6Li}
\end{figure}

In excited-state calculations the quality of the initial wave functions can have a substantial effect on the equilibration time~\cite{Blunt2015_2}. Instead of starting each state from a single determinant, we first perform a small preliminary FCIQMC calculation with $N_\mathrm{w}=10^4$ and use the saved walker distributions to initialize the subsequent large-walker calculations. This choice does not change the converged spectrum, but it provides better multi-configurational initial states and substantially reduces the imaginary-time required to reach stable plateaus.

Figure~\ref{fig:evolution_6Li} shows the resulting imaginary-time evolution for the $N_\mathrm{w}=10^6$ calculation. The five projected energy estimates stabilize rapidly, and the projected $\hat{J}^2$ estimates fluctuate around the expected values $J(J+1)$, identifying the sequence of states as $1^+$, $3^+$, $0^+$, $2_1^+$ and $2_2^+$. This demonstrates that the simultaneous propagation and orthogonalization of several walker ensembles can stably resolve multiple states within the same calculation, without requiring an explicit diagonalization of the full many-body Hamiltonian.

\begin{figure}[t]
    \centering
    \includegraphics[width=0.48\textwidth]{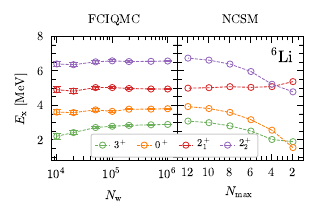}
    \caption{Convergence of the $^6$Li excitation energies $E_\mathrm{x}$ for the lowest four excited states. The left panel shows AS-FCIQMC results as a function of walker number in the $e_\mathrm{max}=6$ model space, and the right panel shows NCSM results as a function of $N_\mathrm{max}$.}
    \label{fig:comparison_6Li}
\end{figure}

The convergence of the excitation energies is shown in Fig.~\ref{fig:comparison_6Li}. In the FCIQMC calculations, the excitation spectrum is already ordered correctly at the smallest walker population, $N_\mathrm{w}=10^4$, and changes only modestly as the walker number is increased to $10^6$. The excitation energies are essentially converged on the scale of the figure for $N_\mathrm{w}\gtrsim 10^5$. In contrast, the NCSM spectrum exhibits a more pronounced dependence on the many-body truncation. At small $N_\mathrm{max}$, the ordering of the low-lying excited states is not yet correct, and the spectrum converges more slowly. The largest NCSM calculation is close to the large-walker FCIQMC result, providing a benchmark of the stochastic excited-state calculation.

This comparison suggests that, for low-lying spectra, FCIQMC can exploit the walker distribution to sample the important regions of the configuration space more flexibly than a global $N_\mathrm{max}$ truncation. While the present $^6$Li example is intended mainly as a demonstration, it indicates that excited-state FCIQMC may become a useful tool for future \textit{ab initio} nuclear spectroscopy in spaces where deterministic diagonalization becomes impractical.

\section{Summary}
This work provides the algorithmic and numerical foundation for the FCIQMC calculations of finite nuclei reported in the companion Letter~\cite{companion_prl}. We formulated FCIQMC for \textit{ab initio} nuclear structure calculations in the harmonic-oscillator single-particle basis and in the corresponding FCI space of Slater determinants. We described the signed-walker imaginary-time propagation, excitation generation, observable estimators, initiator approximation with adaptive shift correction, replica RDM sampling, and the orthogonalized multi-state extension. These ingredients allow high-order correlations to be sampled in the full configuration space without explicitly storing the many-body Hamiltonian matrix or imposing an excitation-rank truncation on the wave function.

The method was validated in the $e_\mathrm{max}=2$ model space for $^4$He, $^8$Be, $^{12}$C and $^{16}$O, where deterministic FCI calculations are available. The FCIQMC energies and RDM-based squared point-proton radii reproduce the FCI results within statistical uncertainties, and the initiator bias is systematically reduced by increasing the total walker number. In the larger $e_\mathrm{max}=10$ space, $^4$He calculation can be converged directly, while for $^8$Be, $^{12}$C and $^{16}$O the slower convergence of RDM-based pure estimators is treated with an energy-constrained power-law extrapolation in $N_\mathrm{w}^{-\alpha}$. The agreement of extrapolations obtained with different adaptive-shift offsets provides an additional check on the residual finite-walker uncertainty, which underlies the energy and charge-radius uncertainties reported in the companion Letter~\cite{companion_prl}. The similar large-walker behavior found in symmetric nuclear matter gives further support to this extrapolation strategy.

We also demonstrated the multi-state FCIQMC algorithm for the five lowest states of $^6$Li in the $e_\mathrm{max}=6$ model space, showing that the method can be extended beyond ground-state calculations. Natural next steps include explicit three-nucleon interactions, larger single-particle basis, more systematic studies of open-shell and clustered nuclei, and applications to collective and electroweak observables that are sensitive to high-order many-body correlations.

\begin{acknowledgments}
We thank S. L. Jin, J. H. Hou, Z. Y. Meng, J. G. Li, J. D. Holt and F. Marino for useful discussions. We thank Gaute Hagen for providing us with the coupled-cluster code for CCSD and CCSDT-3 calculations. 
The NCSM, MBPT and IMSRG calculations were performed using the \texttt{Bigstick}~\cite{code-bigstick}, \texttt{HartreeFock}~\cite{code-mbpt} and \texttt{imsrg++}~\cite{code-imsrg} codes, respectively.
This work was supported by the National Key R\&D Program of China under Grants Nos. 2024YFA1610900 and 2023YFA1606401; the National Natural Science Foundation of China under Grants Nos. 12335007 and 12535008; the Fundamental Research Funds for the Central Universities, Peking University. Numerical calculations in this work were performed at the Huabei Advanced Computing Center and the High-Performance Computing Platform of Peking University.
\end{acknowledgments}

\section*{Data Availability}
The data that support the findings of this article are openly available~\cite{data}.

\bibliographystyle{modified-apsrev4-2.bst}
\bibliography{reference}

@misc{companion_prl,
  title={Full Configuration Interaction Quantum Monte Carlo for Accurate \textit{Ab Initio} Nuclear Structure Calculations}, 
  author={R. Z. Hu and F. R. Xu and B. S. Hu and A. Alavi},
  year={2026},
  eprint={2606.22341},
  archivePrefix={arXiv},
  primaryClass={nucl-th},
  url={https://arxiv.org/abs/2606.22341}, 
}

@misc{code-imsrg,
  author       = {S. R. Stroberg},
  title        = {\texttt{imsrg++}},
  year         = {2026},
  howpublished = {\url{https://github.com/ragnarstroberg/imsrg}},
  url          = {https://github.com/ragnarstroberg/imsrg}
}

@misc{code-mbpt,
  author       = {Takayuki Miyagi},
  title        = {\texttt{HartreeFock}},
  year         = {2026},
  howpublished = {\url{https://github.com/Takayuki-Miyagi/HartreeFock}},
  url          = {https://github.com/Takayuki-Miyagi/HartreeFock}
}

@misc{code-bigstick,
  title         = {BIGSTICK: A flexible configuration-interaction shell-model code (updated)},
  author        = {Calvin W. Johnson and W. Erich Ormand and Kenneth S. McElvain and Ryan Zbikowski and Hongzhang Shan},
  year          = {2025},
  eprint        = {1801.08432},
  archiveprefix = {arXiv},
  primaryclass  = {physics.comp-ph},
  url           = {https://arxiv.org/abs/1801.08432}
}

@misc{code-pyblock,
  author       = {James Spencer},
  title        = {\texttt{pyblock}},
  year         = {2024},
  howpublished = {\url{https://github.com/jsspencer/pyblock}},
  url          = {https://github.com/jsspencer/pyblock}
}

@misc{data,
  howpublished = {\url{https://github.com/hhurongzhe/Data_fciqmc}}
}

@article{sun2025,
  title = {Multiscale Physics of Atomic Nuclei from First Principles},
  author = {Sun, Z. H. and Ekstr\"om, A. and Forss\'en, C. and Hagen, G. and Jansen, G. R. and Papenbrock, T.},
  journal = {Phys. Rev. X},
  volume = {15},
  issue = {1},
  pages = {011028},
  numpages = {22},
  year = {2025},
  month = {Feb},
  publisher = {American Physical Society},
  doi = {10.1103/PhysRevX.15.011028},
  url = {https://link.aps.org/doi/10.1103/PhysRevX.15.011028}
}

@article{Epelbaum2009,
  title     = {Modern theory of nuclear forces},
  author    = {Epelbaum, E. and Hammer, H.-W. and Mei\ss{}ner, Ulf-G.},
  journal   = {Rev. Mod. Phys.},
  volume    = {81},
  issue     = {4},
  pages     = {1773--1825},
  numpages  = {0},
  year      = {2009},
  month     = {Dec},
  publisher = {American Physical Society},
  doi       = {10.1103/RevModPhys.81.1773},
  url       = {https://link.aps.org/doi/10.1103/RevModPhys.81.1773}
}

@article{Hammer2020,
  title     = {Nuclear effective field theory: Status and perspectives},
  author    = {Hammer, H.-W. and K\"onig, Sebastian and van Kolck, U.},
  journal   = {Rev. Mod. Phys.},
  volume    = {92},
  issue     = {2},
  pages     = {025004},
  numpages  = {66},
  year      = {2020},
  month     = {Jun},
  publisher = {American Physical Society},
  doi       = {10.1103/RevModPhys.92.025004},
  url       = {https://link.aps.org/doi/10.1103/RevModPhys.92.025004}
}

@article{Tews2022,
  author  = {Tews, Ingo and Davoudi, Zohreh and Ekstr{\"o}m, Andreas and Holt, Jason D. and Becker, Kevin and Brice{\~n}o, Ra{\'u}l and Dean, David J. and Detmold, William and Drischler, Christian and Duguet, Thomas and Epelbaum, Evgeny and Gasparyan, Ashot and Gegelia, Jambul and Green, Jeremy R. and Grie{\ss}hammer, Harald W. and Hanlon, Andrew D. and Heinz, Matthias and Hergert, Heiko and Hoferichter, Martin and Illa, Marc and Kekejian, David and Kievsky, Alejandro and K{\"o}nig, Sebastian and Krebs, Hermann and Launey, Kristina D. and Lee, Dean and Navr{\'a}til, Petr and Nicholson, Amy and Parre{\~n}o, Assumpta and Phillips, Daniel R. and P{\l}oszajczak, Marek and Ren, Xiu-Lei and Richardson, Thomas R. and Robin, Caroline and Sargsyan, Grigor H. and Savage, Martin J. and Schindler, Matthias R. and Shanahan, Phiala E. and Springer, Roxanne P. and Tichai, Alexander and Kolck, Ubirajara van and Wagman, Michael L. and Walker-Loud, Andr{\'e} and Yang, Chieh-Jen and Zhang, Xilin},
  date    = {2022/09/10},
  doi     = {10.1007/s00601-022-01749-x},
  id      = {Tews2022},
  isbn    = {1432-5411},
  journal = {Few-Body Syst.},
  number  = {4},
  pages   = {67},
  title   = {Nuclear Forces for Precision Nuclear Physics: A Collection of Perspectives},
  url     = {https://doi.org/10.1007/s00601-022-01749-x},
  volume  = {63},
  year    = {2022}
}

@article{Hergert2020,
  title     = {A Guided Tour of {\textit{ab initio}} Nuclear Many-Body Theory},
  volume    = {8},
  issn      = {2296-424X},
  url       = {http://dx.doi.org/10.3389/fphy.2020.00379},
  doi       = {10.3389/fphy.2020.00379},
  journal   = {Front. Phys.},
  publisher = {Frontiers Media SA},
  author    = {Hergert,  Heiko},
  year      = {2020},
  month     = {Oct},
  pages     = {379}
}

@article{Machleidt2011,
  title     = {Chiral effective field theory and nuclear forces},
  volume    = {503},
  issn      = {0370-1573},
  url       = {http://dx.doi.org/10.1016/j.physrep.2011.02.001},
  doi       = {10.1016/j.physrep.2011.02.001},
  number    = {1},
  journal   = {Phys. Rep.},
  publisher = {Elsevier BV},
  author    = {Machleidt,  R. and Entem,  D. R.},
  year      = {2011},
  month     = {June},
  pages     = {1--75}
}

@article{PhysRevLett.103.102502,
  title     = {Three-Nucleon Low-Energy Constants from the Consistency of Interactions and Currents in Chiral Effective Field Theory},
  author    = {Gazit, Doron and Quaglioni, Sofia and Navr\'atil, Petr},
  journal   = {Phys. Rev. Lett.},
  volume    = {103},
  issue     = {10},
  pages     = {102502},
  numpages  = {4},
  year      = {2009},
  month     = {Sep},
  publisher = {American Physical Society},
  doi       = {10.1103/PhysRevLett.103.102502},
  url       = {https://link.aps.org/doi/10.1103/PhysRevLett.103.102502}
}

@article{PhysRevLett.132.232503,
  title     = {Impact of Two-Body Currents on Magnetic Dipole Moments of Nuclei},
  author    = {Miyagi, T. and Cao, X. and Seutin, R. and Bacca, S. and Ruiz, R. F. Garcia and Hebeler, K. and Holt, J. D. and Schwenk, A.},
  journal   = {Phys. Rev. Lett.},
  volume    = {132},
  issue     = {23},
  pages     = {232503},
  numpages  = {6},
  year      = {2024},
  month     = {Jun},
  publisher = {American Physical Society},
  doi       = {10.1103/PhysRevLett.132.232503},
  url       = {https://link.aps.org/doi/10.1103/PhysRevLett.132.232503}
}

@article{PhysRevLett.132.232504,
  title     = {Magnetic Dipole Transition in $^{48}\mathrm{Ca}$},
  author    = {Acharya, B. and Hu, B. S. and Bacca, S. and Hagen, G. and Navr\'atil, P. and Papenbrock, T.},
  journal   = {Phys. Rev. Lett.},
  volume    = {132},
  issue     = {23},
  pages     = {232504},
  numpages  = {7},
  year      = {2024},
  month     = {Jun},
  publisher = {American Physical Society},
  doi       = {10.1103/PhysRevLett.132.232504},
  url       = {https://link.aps.org/doi/10.1103/PhysRevLett.132.232504}
}

@article{PhysRevLett.107.062501,
  title     = {Chiral Two-Body Currents in Nuclei: Gamow-Teller Transitions and Neutrinoless Double-Beta Decay},
  author    = {Men\'endez, J. and Gazit, D. and Schwenk, A.},
  journal   = {Phys. Rev. Lett.},
  volume    = {107},
  issue     = {6},
  pages     = {062501},
  numpages  = {5},
  year      = {2011},
  month     = {Aug},
  publisher = {American Physical Society},
  doi       = {10.1103/PhysRevLett.107.062501},
  url       = {https://link.aps.org/doi/10.1103/PhysRevLett.107.062501}
}

@article{PhysRevC.78.064002,
  title     = {Electromagnetic two-body currents of one- and two-pion range},
  author    = {Pastore, S. and Schiavilla, R. and Goity, J. L.},
  journal   = {Phys. Rev. C},
  volume    = {78},
  issue     = {6},
  pages     = {064002},
  numpages  = {25},
  year      = {2008},
  month     = {Dec},
  publisher = {American Physical Society},
  doi       = {10.1103/PhysRevC.78.064002},
  url       = {https://link.aps.org/doi/10.1103/PhysRevC.78.064002}
}

@article{PhysRevC.80.034004,
  title     = {Electromagnetic currents and magnetic moments in chiral effective field theory $(\ensuremath{\chi}\mathrm{EFT})$},
  author    = {Pastore, S. and Girlanda, L. and Schiavilla, R. and Viviani, M. and Wiringa, R. B.},
  journal   = {Phys. Rev. C},
  volume    = {80},
  issue     = {3},
  pages     = {034004},
  numpages  = {22},
  year      = {2009},
  month     = {Sep},
  publisher = {American Physical Society},
  doi       = {10.1103/PhysRevC.80.034004},
  url       = {https://link.aps.org/doi/10.1103/PhysRevC.80.034004}
}

@article{PhysRevC.84.024001,
  title     = {Two-nucleon electromagnetic charge operator in chiral effective field theory ($\ensuremath{\chi}$EFT) up to one loop},
  author    = {Pastore, S. and Girlanda, L. and Schiavilla, R. and Viviani, M.},
  journal   = {Phys. Rev. C},
  volume    = {84},
  issue     = {2},
  pages     = {024001},
  numpages  = {15},
  year      = {2011},
  month     = {Aug},
  publisher = {American Physical Society},
  doi       = {10.1103/PhysRevC.84.024001},
  url       = {https://link.aps.org/doi/10.1103/PhysRevC.84.024001}
}

@article{PhysRevC.102.025501,
  title     = {Chiral effective field theory calculations of weak transitions in light nuclei},
  author    = {King, G. B. and Andreoli, L. and Pastore, S. and Piarulli, M. and Schiavilla, R. and Wiringa, R. B. and Carlson, J. and Gandolfi, S.},
  journal   = {Phys. Rev. C},
  volume    = {102},
  issue     = {2},
  pages     = {025501},
  numpages  = {13},
  year      = {2020},
  month     = {Aug},
  publisher = {American Physical Society},
  doi       = {10.1103/PhysRevC.102.025501},
  url       = {https://link.aps.org/doi/10.1103/PhysRevC.102.025501}
}

@article{Krebs2020,
  title     = {Nuclear currents in chiral effective field theory},
  volume    = {56},
  issn      = {1434-601X},
  url       = {http://dx.doi.org/10.1140/epja/s10050-020-00230-9},
  doi       = {10.1140/epja/s10050-020-00230-9},
  number    = {9},
  journal   = {Eur. Phys. J. A},
  publisher = {Springer Science and Business Media LLC},
  author    = {Krebs,  Hermann},
  year      = {2020},
  month     = {Sept},
  pages     = {234}
}

@article{Gysbers2019,
  title     = {Discrepancy between experimental and theoretical $\beta$-decay rates resolved from first principles},
  volume    = {15},
  issn      = {1745-2481},
  url       = {http://dx.doi.org/10.1038/s41567-019-0450-7},
  doi       = {10.1038/s41567-019-0450-7},
  number    = {5},
  journal   = {Nat. Phys.},
  publisher = {Springer Science and Business Media LLC},
  author    = {Gysbers,  P. and Hagen,  G. and Holt,  J. D. and Jansen,  G. R. and Morris,  T. D. and Navr\'atil,  P. and Papenbrock,  T. and Quaglioni,  S. and Schwenk,  A. and Stroberg,  S. R. and Wendt,  K. A.},
  year      = {2019},
  month     = {Mar},
  pages     = {428--431}
}

@article{PhysRevLett.84.5728,
  title     = {Properties of ${}^{12}$C in the {\textit{Ab Initio}} Nuclear Shell Model},
  author    = {Navr\'atil, P. and Vary, J. P. and Barrett, B. R.},
  journal   = {Phys. Rev. Lett.},
  volume    = {84},
  issue     = {25},
  pages     = {5728--5731},
  numpages  = {0},
  year      = {2000},
  month     = {Jun},
  publisher = {American Physical Society},
  doi       = {10.1103/PhysRevLett.84.5728},
  url       = {https://link.aps.org/doi/10.1103/PhysRevLett.84.5728}
}

@article{Barrett2013,
  title     = {{\textit{Ab initio}} no core shell model},
  volume    = {69},
  issn      = {0146-6410},
  url       = {http://dx.doi.org/10.1016/j.ppnp.2012.10.003},
  doi       = {10.1016/j.ppnp.2012.10.003},
  journal   = {Prog. Part. Nucl. Phys.},
  publisher = {Elsevier BV},
  author    = {Barrett,  Bruce R. and Navr\'atil,  Petr and Vary,  James P.},
  year      = {2013},
  month     = {Mar},
  pages     = {131--181}
}

@article{c3st-tp13,
  title     = {Unexpected Rise in Nuclear Collectivity from Short-Range Physics},
  author    = {Becker, Kevin S. and Launey, Kristina D. and Ekstr\"om, Andreas and Tomas Dytrych and Langr, Daniel and Sargsyan, Grigor H. and Draayer, Jerry P.},
  journal   = {Phys. Rev. Lett.},
  volume    = {136},
  issue     = {2},
  pages     = {022501},
  numpages  = {8},
  year      = {2026},
  month     = {Jan},
  publisher = {American Physical Society},
  doi       = {10.1103/c3st-tp13},
  url       = {https://link.aps.org/doi/10.1103/c3st-tp13}
}

@article{ROTH2010272,
  title    = {Pad{\'e}-Resummed High-Order Perturbation Theory for Nuclear Structure Calculations},
  journal  = {Phys. Lett. B},
  volume   = {683},
  number   = {4},
  pages    = {272--277},
  year     = {2010},
  issn     = {0370-2693},
  doi      = {https://doi.org/10.1016/j.physletb.2009.12.046},
  url      = {https://www.sciencedirect.com/science/article/pii/S037026930901507X},
  author   = {Robert Roth and Joachim Langhammer}
}

@article{PhysRevC.94.014303,
  title     = {{\textit{Ab initio}} Nuclear Many-Body Perturbation Calculations in the Hartree-Fock Basis},
  author    = {Hu, B. S. and Xu, F. R. and Sun, Z. H. and Vary, J. P. and Li, T.},
  journal   = {Phys. Rev. C},
  volume    = {94},
  issue     = {1},
  pages     = {014303},
  numpages  = {11},
  year      = {2016},
  month     = {Jul},
  publisher = {American Physical Society},
  doi       = {10.1103/PhysRevC.94.014303},
  url       = {https://link.aps.org/doi/10.1103/PhysRevC.94.014303}
}

@article{TICHAI2018448,
  title    = {Open-Shell Nuclei from No-Core Shell Model with Perturbative Improvement},
  journal  = {Phys. Lett. B},
  volume   = {786},
  pages    = {448--452},
  year     = {2018},
  issn     = {0370-2693},
  doi      = {https://doi.org/10.1016/j.physletb.2018.10.029},
  url      = {https://www.sciencedirect.com/science/article/pii/S0370269318307986},
  author   = {Alexander Tichai and Eskendr Gebrerufael and Klaus Vobig and Robert Roth}
}

@article{PhysRevLett.122.042501,
  title     = {Chiral Interactions Up to Next-To-Next-To-Next-To-Leading Order and Nuclear Saturation},
  author    = {Drischler, C. and Hebeler, K. and Schwenk, A.},
  journal   = {Phys. Rev. Lett.},
  volume    = {122},
  issue     = {4},
  pages     = {042501},
  numpages  = {6},
  year      = {2019},
  month     = {Jan},
  publisher = {American Physical Society},
  doi       = {10.1103/PhysRevLett.122.042501},
  url       = {https://link.aps.org/doi/10.1103/PhysRevLett.122.042501}
}

@article{10.3389/fphy.2020.00164,
  title     = {Many-Body Perturbation Theories for Finite Nuclei},
  volume    = {8},
  issn      = {2296-424X},
  url       = {http://dx.doi.org/10.3389/fphy.2020.00164},
  doi       = {10.3389/fphy.2020.00164},
  journal   = {Front. Phys.},
  publisher = {Frontiers Media SA},
  author    = {Tichai,  Alexander and Roth,  Robert and Duguet,  Thomas},
  year      = {2020},
  month     = {June},
  pages     = {164}
}

@article{PhysRevLett.106.222502,
  title     = {In-Medium Similarity Renormalization Group For Nuclei},
  author    = {Tsukiyama, K. and Bogner, S. K. and Schwenk, A.},
  journal   = {Phys. Rev. Lett.},
  volume    = {106},
  issue     = {22},
  pages     = {222502},
  numpages  = {4},
  year      = {2011},
  month     = {Jun},
  publisher = {American Physical Society},
  doi       = {10.1103/PhysRevLett.106.222502},
  url       = {https://link.aps.org/doi/10.1103/PhysRevLett.106.222502}
}

@article{PhysRevC.85.061304,
  title     = {In-medium similarity renormalization group for open-shell nuclei},
  author    = {Tsukiyama, K. and Bogner, S. K. and Schwenk, A.},
  journal   = {Phys. Rev. C},
  volume    = {85},
  issue     = {6},
  pages     = {061304},
  numpages  = {4},
  year      = {2012},
  month     = {Jun},
  publisher = {American Physical Society},
  doi       = {10.1103/PhysRevC.85.061304},
  url       = {https://link.aps.org/doi/10.1103/PhysRevC.85.061304}
}

@article{Hergert2016,
  title     = {The In-Medium Similarity Renormalization Group: A novel {\textit{ab initio}} method for nuclei},
  volume    = {621},
  issn      = {0370-1573},
  url       = {http://dx.doi.org/10.1016/j.physrep.2015.12.007},
  doi       = {10.1016/j.physrep.2015.12.007},
  journal   = {Phys. Rep.},
  publisher = {Elsevier BV},
  author    = {Hergert,  H. and Bogner,  S. K. and Morris,  T. D. and Schwenk,  A. and Tsukiyama,  K.},
  year      = {2016},
  month     = {Mar},
  pages     = {165--222}
}

@article{PhysRevC.92.034331,
  title     = {Magnus expansion and in-medium similarity renormalization group},
  author    = {Morris, T. D. and Parzuchowski, N. M. and Bogner, S. K.},
  journal   = {Phys. Rev. C},
  volume    = {92},
  issue     = {3},
  pages     = {034331},
  numpages  = {12},
  year      = {2015},
  month     = {Sep},
  publisher = {American Physical Society},
  doi       = {10.1103/PhysRevC.92.034331},
  url       = {https://link.aps.org/doi/10.1103/PhysRevC.92.034331}
}

@article{PhysRevC.103.044318,
  title     = {In-medium similarity renormalization group with three-body operators},
  author    = {Heinz, M. and Tichai, A. and Hoppe, J. and Hebeler, K. and Schwenk, A.},
  journal   = {Phys. Rev. C},
  volume    = {103},
  issue     = {4},
  pages     = {044318},
  numpages  = {21},
  year      = {2021},
  month     = {Apr},
  publisher = {American Physical Society},
  doi       = {10.1103/PhysRevC.103.044318},
  url       = {https://link.aps.org/doi/10.1103/PhysRevC.103.044318}
}

@article{PhysRevLett.109.252501,
  title     = {Structure and Rotations of the Hoyle State},
  author    = {Epelbaum, Evgeny and Krebs, Hermann and L\"ahde, Timo A. and Lee, Dean and Mei\ss{}ner, Ulf-G.},
  journal   = {Phys. Rev. Lett.},
  volume    = {109},
  issue     = {25},
  pages     = {252501},
  numpages  = {4},
  year      = {2012},
  month     = {Dec},
  publisher = {American Physical Society},
  doi       = {10.1103/PhysRevLett.109.252501},
  url       = {https://link.aps.org/doi/10.1103/PhysRevLett.109.252501}
}

@article{PhysRevC.91.014310,
  title     = {Emergence of rotational bands in {\textit{ab initio}} no-core configuration interaction calculations of the Be isotopes},
  author    = {Maris, P. and Caprio, M. A. and Vary, J. P.},
  journal   = {Phys. Rev. C},
  volume    = {91},
  issue     = {1},
  pages     = {014310},
  numpages  = {29},
  year      = {2015},
  month     = {Jan},
  publisher = {American Physical Society},
  doi       = {10.1103/PhysRevC.91.014310},
  url       = {https://link.aps.org/doi/10.1103/PhysRevC.91.014310}
}

@article{PhysRevLett.112.102501,
  title     = {{\textit{Ab Initio}} Calculation of the Spectrum and Structure of $^{16}\mathrm{O}$},
  author    = {Epelbaum, Evgeny and Krebs, Hermann and L\"ahde, Timo A. and Lee, Dean and Mei\ss{}ner, Ulf-G. and Rupak, Gautam},
  journal   = {Phys. Rev. Lett.},
  volume    = {112},
  issue     = {10},
  pages     = {102501},
  numpages  = {5},
  year      = {2014},
  month     = {Mar},
  publisher = {American Physical Society},
  doi       = {10.1103/PhysRevLett.112.102501},
  url       = {https://link.aps.org/doi/10.1103/PhysRevLett.112.102501}
}

@article{mmy4-3wrp,
  title     = {Chiral interactions and superfluidity in the calcium isotopic chain},
  author    = {Scalesi, A. and Ekstr\"om, A. and Forss\'en, C. and Hagen, G.},
  journal   = {Phys. Rev. C},
  volume    = {113},
  issue     = {5},
  pages     = {L051303},
  numpages  = {6},
  year      = {2026},
  month     = {May},
  publisher = {American Physical Society},
  doi       = {10.1103/mmy4-3wrp},
  url       = {https://link.aps.org/doi/10.1103/mmy4-3wrp}
}

@article{Tichai2024,
  title     = {Towards heavy-mass {\textit{ab initio}} nuclear structure: Open-shell Ca, Ni and Sn isotopes from Bogoliubov coupled-cluster theory},
  volume    = {851},
  issn      = {0370-2693},
  url       = {http://dx.doi.org/10.1016/j.physletb.2024.138571},
  doi       = {10.1016/j.physletb.2024.138571},
  journal   = {Phys. Lett. B},
  publisher = {Elsevier BV},
  author    = {Tichai,  A. and Demol,  P. and Duguet,  T.},
  year      = {2024},
  month     = {Apr},
  pages     = {138571}
}

@article{Tichai2018_2,
  title     = {Bogoliubov many-body perturbation theory for open-shell nuclei},
  volume    = {786},
  issn      = {0370-2693},
  url       = {http://dx.doi.org/10.1016/j.physletb.2018.09.044},
  doi       = {10.1016/j.physletb.2018.09.044},
  journal   = {Phys. Lett. B},
  publisher = {Elsevier BV},
  author    = {Tichai,  A. and Arthuis,  P. and Duguet,  T. and Hergert,  H. and Som\`a,  V. and Roth,  R.},
  year      = {2018},
  month     = {Nov},
  pages     = {195--200}
}

@article{Otsuka2022,
  title     = {$\alpha$-Clustering in atomic nuclei from first principles with statistical learning and the Hoyle state character},
  volume    = {13},
  issn      = {2041-1723},
  url       = {http://dx.doi.org/10.1038/s41467-022-29582-0},
  doi       = {10.1038/s41467-022-29582-0},
  number    = {1},
  pages={2234},
  journal   = {Nat. Commun.},
  publisher = {Springer Science and Business Media LLC},
  author    = {Otsuka,  T. and Abe,  T. and Yoshida,  T. and Tsunoda,  Y. and Shimizu,  N. and Itagaki,  N. and Utsuno,  Y. and Vary,  J. and Maris,  P. and Ueno,  H.},
  year      = {2022},
  month     = {Apr}
}

@article{PhysRevLett.124.232501,
  title     = {{\textit{Ab Initio}} Treatment of Collective Correlations and the Neutrinoless Double Beta Decay of $^{48}\mathrm{Ca}$},
  author    = {Yao, J. M. and Bally, B. and Engel, J. and Wirth, R. and Rodr\'{\i}guez, T. R. and Hergert, H.},
  journal   = {Phys. Rev. Lett.},
  volume    = {124},
  issue     = {23},
  pages     = {232501},
  numpages  = {6},
  year      = {2020},
  month     = {Jun},
  publisher = {American Physical Society},
  doi       = {10.1103/PhysRevLett.124.232501},
  url       = {https://link.aps.org/doi/10.1103/PhysRevLett.124.232501}
}

@article{PhysRevC.105.L061303,
  title     = {Deformed in-medium similarity renormalization group},
  author    = {Yuan, Q. and Fan, S. Q. and Hu, B. S. and Li, J. G. and Zhang, S. and Wang, S. M. and Sun, Z. H. and Ma, Y. Z. and Xu, F. R.},
  journal   = {Phys. Rev. C},
  volume    = {105},
  issue     = {6},
  pages     = {L061303},
  numpages  = {7},
  year      = {2022},
  month     = {Jun},
  publisher = {American Physical Society},
  doi       = {10.1103/PhysRevC.105.L061303},
  url       = {https://link.aps.org/doi/10.1103/PhysRevC.105.L061303}
}

@article{PhysRevC.99.061302,
  title     = {{\textit{Ab initio}} Gamow in-medium similarity renormalization group with resonance and continuum},
  author    = {Hu, B. S. and Wu, Q. and Sun, Z. H. and Xu, F. R.},
  journal   = {Phys. Rev. C},
  volume    = {99},
  issue     = {6},
  pages     = {061302},
  numpages  = {6},
  year      = {2019},
  month     = {Jun},
  publisher = {American Physical Society},
  doi       = {10.1103/PhysRevC.99.061302},
  url       = {https://link.aps.org/doi/10.1103/PhysRevC.99.061302}
}

@article{PhysRevLett.103.082501,
  title     = {Evolution of Nuclear Many-Body Forces with the Similarity Renormalization Group},
  author    = {Jurgenson, E. D. and Navr\'atil, P. and Furnstahl, R. J.},
  journal   = {Phys. Rev. Lett.},
  volume    = {103},
  issue     = {8},
  pages     = {082501},
  numpages  = {4},
  year      = {2009},
  month     = {Aug},
  publisher = {American Physical Society},
  doi       = {10.1103/PhysRevLett.103.082501},
  url       = {https://link.aps.org/doi/10.1103/PhysRevLett.103.082501}
}

@article{Zhen2025,
  title     = {Non-perturbative calculations of nuclear matter using in-medium similarity renormalization group},
  volume    = {862},
  issn      = {0370-2693},
  url       = {http://dx.doi.org/10.1016/j.physletb.2025.139350},
  doi       = {10.1016/j.physletb.2025.139350},
  journal   = {Phys. Lett. B},
  publisher = {Elsevier BV},
  author    = {Zhen,  Xin and Hu,  Rongzhe and Shang,  Haoyu and Chen,  Jiawei and Pei, J. C. and Xu, F. R.},
  year      = {2025},
  month     = Mar,
  pages     = {139350}
}

@article{PhysRevC.111.034311,
  title     = {Improved structure of calcium isotopes from {\textit{ab initio}} calculations},
  author    = {Heinz, M. and Miyagi, T. and Stroberg, S. R. and Tichai, A. and Hebeler, K. and Schwenk, A.},
  journal   = {Phys. Rev. C},
  volume    = {111},
  issue     = {3},
  pages     = {034311},
  numpages  = {14},
  year      = {2025},
  month     = {Mar},
  publisher = {American Physical Society},
  doi       = {10.1103/PhysRevC.111.034311},
  url       = {https://link.aps.org/doi/10.1103/PhysRevC.111.034311}
}

@article{PhysRevC.110.044316,
  title     = {In-medium similarity renormalization group with flowing 3-body operators, and approximations thereof},
  author    = {Stroberg, S. R. and Morris, T. D. and He, B. C.},
  journal   = {Phys. Rev. C},
  volume    = {110},
  issue     = {4},
  pages     = {044316},
  numpages  = {16},
  year      = {2024},
  month     = {Oct},
  publisher = {American Physical Society},
  doi       = {10.1103/PhysRevC.110.044316},
  url       = {https://link.aps.org/doi/10.1103/PhysRevC.110.044316}
}

@article{PhysRevC.110.044317,
  title     = {Factorized approximation to the in-medium similarity renormalization group IMSRG(3)},
  author    = {He, B. C. and Stroberg, S. R.},
  journal   = {Phys. Rev. C},
  volume    = {110},
  issue     = {4},
  pages     = {044317},
  numpages  = {16},
  year      = {2024},
  month     = {Oct},
  publisher = {American Physical Society},
  doi       = {10.1103/PhysRevC.110.044317},
  url       = {https://link.aps.org/doi/10.1103/PhysRevC.110.044317}
}

@article{PhysRevLett.113.142502,
  title     = {{\textit{Ab Initio}} Coupled-Cluster Effective Interactions for the Shell Model: Application to Neutron-Rich Oxygen and Carbon Isotopes},
  author    = {Jansen, G. R. and Engel, J. and Hagen, G. and Navratil, P. and Signoracci, A.},
  journal   = {Phys. Rev. Lett.},
  volume    = {113},
  issue     = {14},
  pages     = {142502},
  numpages  = {5},
  year      = {2014},
  month     = {Oct},
  publisher = {American Physical Society},
  doi       = {10.1103/PhysRevLett.113.142502},
  url       = {https://link.aps.org/doi/10.1103/PhysRevLett.113.142502}
}

@article{PhysRevC.76.034302,
  title     = {Coupled-cluster theory for three-body Hamiltonians},
  author    = {Hagen, G. and Papenbrock, T. and Dean, D. J. and Schwenk, A. and Nogga, A. and W\l{}och, M. and Piecuch, P.},
  journal   = {Phys. Rev. C},
  volume    = {76},
  issue     = {3},
  pages     = {034302},
  numpages  = {11},
  year      = {2007},
  month     = {Sep},
  publisher = {American Physical Society},
  doi       = {10.1103/PhysRevC.76.034302},
  url       = {https://link.aps.org/doi/10.1103/PhysRevC.76.034302}
}

@article{RevModPhys.79.291,
  title     = {Coupled-cluster theory in quantum chemistry},
  author    = {Bartlett, Rodney J. and Musia\l{}, Monika},
  journal   = {Rev. Mod. Phys.},
  volume    = {79},
  issue     = {1},
  pages     = {291--352},
  numpages  = {0},
  year      = {2007},
  month     = {Feb},
  publisher = {American Physical Society},
  doi       = {10.1103/RevModPhys.79.291},
  url       = {https://link.aps.org/doi/10.1103/RevModPhys.79.291}
}

@article{PhysRevLett.92.132501,
  title     = {Coupled Cluster Calculations of Ground and Excited States of Nuclei},
  author    = {Kowalski, K. and Dean, D. J. and Hjorth-Jensen, M. and Papenbrock, T. and Piecuch, P.},
  journal   = {Phys. Rev. Lett.},
  volume    = {92},
  issue     = {13},
  pages     = {132501},
  numpages  = {4},
  year      = {2004},
  month     = {Apr},
  publisher = {American Physical Society},
  doi       = {10.1103/PhysRevLett.92.132501},
  url       = {https://link.aps.org/doi/10.1103/PhysRevLett.92.132501}
}

@article{PhysRevC.82.034330,
  title     = {{\textit{Ab initio}} coupled-cluster approach to nuclear structure with modern nucleon-nucleon interactions},
  author    = {Hagen, G. and Papenbrock, T. and Dean, D. J. and Hjorth-Jensen, M.},
  journal   = {Phys. Rev. C},
  volume    = {82},
  issue     = {3},
  pages     = {034330},
  numpages  = {22},
  year      = {2010},
  month     = {Sep},
  publisher = {American Physical Society},
  doi       = {10.1103/PhysRevC.82.034330},
  url       = {https://link.aps.org/doi/10.1103/PhysRevC.82.034330}
}

@article{PhysRevC.83.054306,
  title     = {Toward open-shell nuclei with coupled-cluster theory},
  author    = {Jansen, G. R. and Hjorth-Jensen, M. and Hagen, G. and Papenbrock, T.},
  journal   = {Phys. Rev. C},
  volume    = {83},
  issue     = {5},
  pages     = {054306},
  numpages  = {9},
  year      = {2011},
  month     = {May},
  publisher = {American Physical Society},
  doi       = {10.1103/PhysRevC.83.054306},
  url       = {https://link.aps.org/doi/10.1103/PhysRevC.83.054306}
}

@article{Hagen2014,
  title     = {Coupled-cluster computations of atomic nuclei},
  volume    = {77},
  issn      = {1361-6633},
  url       = {http://dx.doi.org/10.1088/0034-4885/77/9/096302},
  doi       = {10.1088/0034-4885/77/9/096302},
  number    = {9},
  journal   = {Rep. Prog. Phys.},
  publisher = {IOP Publishing},
  author    = {Hagen, G. and Papenbrock, T. and Hjorth-Jensen, M. and Dean, D. J.},
  year      = {2014},
  month     = {Sept},
  pages     = {096302}
}

@article{p297-y8vq,
  title     = {From closed shells to open shells: Coupled-cluster calculations of atomic nuclei},
  author    = {Marino, F. and Bonaiti, F. and Demol, P. and Bacca, S. and Duguet, T. and Hagen, G. and Jansen, G. R. and Papenbrock, T. and Tichai, A.},
  journal   = {Phys. Rev. C},
  volume    = {113},
  issue     = {4},
  pages     = {044301},
  numpages  = {13},
  year      = {2026},
  month     = {Apr},
  publisher = {American Physical Society},
  doi       = {10.1103/p297-y8vq},
  url       = {https://link.aps.org/doi/10.1103/p297-y8vq}
}

@article{Dickhoff2004,
  title     = {Self-consistent Green's function method for nuclei and nuclear matter},
  volume    = {52},
  issn      = {0146-6410},
  url       = {http://dx.doi.org/10.1016/j.ppnp.2004.02.038},
  doi       = {10.1016/j.ppnp.2004.02.038},
  number    = {2},
  journal   = {Prog. Part. Nucl. Phys.},
  publisher = {Elsevier BV},
  author    = {Dickhoff,  W. H. and Barbieri,  C.},
  year      = {2004},
  month     = {Apr},
  pages     = {377--496}
}

@article{Som2020,
  title     = {Self-Consistent Green's Function Theory for Atomic Nuclei},
  volume    = {8},
  issn      = {2296-424X},
  url       = {http://dx.doi.org/10.3389/fphy.2020.00340},
  doi       = {10.3389/fphy.2020.00340},
  journal   = {Front. Phys.},
  publisher = {Frontiers Media SA},
  author    = {Som\`a,  Vittorio},
  year      = {2020},
  month     = {Sept},
  pages     = {340}
}

@article{PhysRevC.101.014318,
  title     = {Novel chiral Hamiltonian and observables in light and medium-mass nuclei},
  author    = {Som\`a, V. and Navr\'atil, P. and Raimondi, F. and Barbieri, C. and Duguet, T.},
  journal   = {Phys. Rev. C},
  volume    = {101},
  issue     = {1},
  pages     = {014318},
  numpages  = {19},
  year      = {2020},
  month     = {Jan},
  publisher = {American Physical Society},
  doi       = {10.1103/PhysRevC.101.014318},
  url       = {https://link.aps.org/doi/10.1103/PhysRevC.101.014318}
}

@article{PhysRevLett.134.182502,
  title     = {Diagrammatic Monte Carlo for Finite Systems at Zero Temperature},
  author    = {Brolli, Stefano and Barbieri, Carlo and Vigezzi, Enrico},
  journal   = {Phys. Rev. Lett.},
  volume    = {134},
  issue     = {18},
  pages     = {182502},
  numpages  = {6},
  year      = {2025},
  month     = {May},
  publisher = {American Physical Society},
  doi       = {10.1103/PhysRevLett.134.182502},
  url       = {https://link.aps.org/doi/10.1103/PhysRevLett.134.182502}
}

@article{Booth2014,
  title     = {Linear-scaling and parallelisable algorithms for stochastic quantum chemistry},
  volume    = {112},
  issn      = {1362-3028},
  url       = {http://dx.doi.org/10.1080/00268976.2013.877165},
  doi       = {10.1080/00268976.2013.877165},
  number    = {14},
  journal   = {Mol. Phys.},
  publisher = {Informa UK Limited},
  author    = {Booth,  George H. and Smart,  Simon D. and Alavi,  Ali},
  year      = {2014},
  month     = {Jan},
  pages     = {1855--1869}
}

@misc{drischler2026,
  title         = {Many-body perturbation theory for the nuclear equation of state up to fifth order},
  author        = {C. Drischler and K. S. McElvain and P. Arthuis},
  year          = {2026},
  eprint        = {2603.24532},
  archiveprefix = {arXiv},
  primaryclass  = {nucl-th},
  url           = {https://arxiv.org/abs/2603.24532}
}

@article{Booth2009,
  title     = {Fermion Monte Carlo without fixed nodes: A game of life,  death,  and annihilation in Slater determinant space},
  volume    = {131},
  issn      = {1089-7690},
  url       = {http://dx.doi.org/10.1063/1.3193710},
  doi       = {10.1063/1.3193710},
  number    = {5},
  journal   = {J. Chem. Phys.},
  publisher = {AIP Publishing},
  author    = {Booth,  George H. and Thom,  Alex J. W. and Alavi,  Ali},
  year      = {2009},
  month     = {Aug},
  pages = {054106}
}

@phdthesis{Dobrautz2025,
  author = {Werner Dobrautz},
  title  = {Development of Full Configuration Interaction Quantum Monte Carlo Methods for Strongly Correlated Electron Systems},
  school = {University of Stuttgart},
  year   = {2019}
}

@phdthesis{Overy2012,
  author = {Catherine Overy},
  title  = {Reduced Density Matrices and Stochastic Quantum Chemistry},
  school = {University of Cambridge},
  year   = {2012}
}

@phdthesis{Hosseini2024,
  author = {Seyed Mohammadreza Hosseini},
  title  = {Combining Orbital and Real Space Quantum Monte Carlo Methods},
  school = {University of Stuttgart},
  year   = {2024}
}

@article{Dobrautz2019,
  title     = {Efficient formulation of full configuration interaction quantum Monte Carlo in a spin eigenbasis via the graphical unitary group approach},
  volume    = {151},
  issn      = {1089-7690},
  url       = {http://dx.doi.org/10.1063/1.5108908},
  doi       = {10.1063/1.5108908},
  number    = {9},
  journal   = {J. Chem. Phys.},
  publisher = {AIP Publishing},
  author    = {Dobrautz,  Werner and Smart,  Simon D. and Alavi,  Ali},
  year      = {2019},
  month     = {Sept},
  pages = {094104}
}

@article{Booth2010,
  title     = {Approaching chemical accuracy using full configuration-interaction quantum Monte Carlo: A study of ionization potentials},
  volume    = {132},
  issn      = {1089-7690},
  url       = {http://dx.doi.org/10.1063/1.3407895},
  doi       = {10.1063/1.3407895},
  number    = {17},
  journal   = {J. Chem. Phys.},
  publisher = {AIP Publishing},
  author    = {Booth,  George H. and Alavi,  Ali},
  year      = {2010},
  month     = {May},
  pages = {174104}
}

@article{Booth2011,
  title     = {Breaking the carbon dimer: The challenges of multiple bond dissociation with full configuration interaction quantum Monte Carlo methods},
  volume    = {135},
  issn      = {1089-7690},
  url       = {http://dx.doi.org/10.1063/1.3624383},
  doi       = {10.1063/1.3624383},
  number    = {8},
  journal   = {J. Chem. Phys.},
  publisher = {AIP Publishing},
  author    = {Booth,  George H. and Cleland,  Deidre and Thom,  Alex J. W. and Alavi,  Ali},
  year      = {2011},
  month     = {Aug},
  pages = {084104}
}

@article{Cleland2011,
  title     = {A study of electron affinities using the initiator approach to full configuration interaction quantum Monte Carlo},
  volume    = {134},
  issn      = {1089-7690},
  url       = {http://dx.doi.org/10.1063/1.3525712},
  doi       = {10.1063/1.3525712},
  number    = {2},
  journal   = {J. Chem. Phys.},
  publisher = {AIP Publishing},
  author    = {Cleland,  D. M. and Booth,  George H. and Alavi,  Ali},
  year      = {2011},
  month     = {Jan},
  pages = {024112}
}

@article{PhysRevLett.109.230201,
  title     = {Semistochastic Projector Monte Carlo Method},
  author    = {Petruzielo, F. R. and Holmes, A. A. and Changlani, Hitesh J. and Nightingale, M. P. and Umrigar, C. J.},
  journal   = {Phys. Rev. Lett.},
  volume    = {109},
  issue     = {23},
  pages     = {230201},
  numpages  = {5},
  year      = {2012},
  month     = {Dec},
  publisher = {American Physical Society},
  doi       = {10.1103/PhysRevLett.109.230201},
  url       = {https://link.aps.org/doi/10.1103/PhysRevLett.109.230201}
}

@article{Shepherd2012,
  title     = {Investigation of the full configuration interaction quantum Monte Carlo method using homogeneous electron gas models},
  volume    = {136},
  issn      = {1089-7690},
  url       = {http://dx.doi.org/10.1063/1.4720076},
  doi       = {10.1063/1.4720076},
  number    = {24},
  journal   = {J. Chem. Phys.},
  publisher = {AIP Publishing},
  author    = {Shepherd,  James J. and Booth,  George H. and Alavi,  Ali},
  year      = {2012},
  month     = {June},
  pages = {244101}
}

@article{PhysRevB.85.081103,
  title     = {Full configuration interaction perspective on the homogeneous electron gas},
  author    = {Shepherd, James J. and Booth, George and Gr\"uneis, Andreas and Alavi, Ali},
  journal   = {Phys. Rev. B},
  volume    = {85},
  issue     = {8},
  pages     = {081103},
  numpages  = {4},
  year      = {2012},
  month     = {Feb},
  publisher = {American Physical Society},
  doi       = {10.1103/PhysRevB.85.081103},
  url       = {https://link.aps.org/doi/10.1103/PhysRevB.85.081103}
}

@article{Booth2012n,
  title     = {Towards an exact description of electronic wavefunctions in real solids},
  volume    = {493},
  issn      = {1476-4687},
  url       = {http://dx.doi.org/10.1038/nature11770},
  doi       = {10.1038/nature11770},
  number    = {7432},
  journal   = {Nature},
  publisher = {Springer Science and Business Media LLC},
  author    = {Booth,  George H. and Gr\"{u}neis,  Andreas and Kresse,  Georg and Alavi,  Ali},
  year      = {2012},
  month     = {Dec},
  pages     = {365--370}
}

@article{Ghanem2019,
  title     = {Unbiasing the initiator approximation in full configuration interaction quantum Monte Carlo},
  volume    = {151},
  issn      = {1089-7690},
  url       = {http://dx.doi.org/10.1063/1.5134006},
  doi       = {10.1063/1.5134006},
  number    = {22},
  journal   = {J. Chem. Phys.},
  publisher = {AIP Publishing},
  author    = {Ghanem,  Khaldoon and Lozovoi,  Alexander Y. and Alavi,  Ali},
  year      = {2019},
  month     = {Dec},
  pages = {224108}
}

@article{Ghanem2020,
  title     = {The adaptive shift method in full configuration interaction quantum Monte Carlo: Development and applications},
  volume    = {153},
  issn      = {1089-7690},
  url       = {http://dx.doi.org/10.1063/5.0032617},
  doi       = {10.1063/5.0032617},
  number    = {22},
  journal   = {J. Chem. Phys.},
  publisher = {AIP Publishing},
  author    = {Ghanem,  Khaldoon and Guther,  Kai and Alavi,  Ali},
  year      = {2020},
  month     = {Dec},
  pages = {224115}
}

@article{PhysRevB.90.155130,
  title     = {Sign problem in full configuration interaction quantum Monte Carlo: Linear and sublinear representation regimes for the exact wave function},
  author    = {Shepherd, James J. and Scuseria, Gustavo E. and Spencer, James S.},
  journal   = {Phys. Rev. B},
  volume    = {90},
  issue     = {15},
  pages     = {155130},
  numpages  = {5},
  year      = {2014},
  month     = {Oct},
  publisher = {American Physical Society},
  doi       = {10.1103/PhysRevB.90.155130},
  url       = {https://link.aps.org/doi/10.1103/PhysRevB.90.155130}
}

@article{Blunt2015_2,
  title     = {An excited-state approach within full configuration interaction quantum Monte Carlo},
  volume    = {143},
  issn      = {1089-7690},
  url       = {http://dx.doi.org/10.1063/1.4932595},
  doi       = {10.1063/1.4932595},
  number    = {13},
  journal   = {J. Chem. Phys.},
  publisher = {AIP Publishing},
  author    = {Blunt,  N. S. and Smart,  Simon D. and Booth,  George H. and Alavi,  Ali},
  year      = {2015},
  month     = {Oct},
  pages = {134117}
}

@article{Blunt2017,
  title     = {Density matrices in full configuration interaction quantum Monte Carlo: Excited states,  transition dipole moments,  and parallel distribution},
  volume    = {146},
  issn      = {1089-7690},
  url       = {http://dx.doi.org/10.1063/1.4986963},
  doi       = {10.1063/1.4986963},
  number    = {24},
  journal   = {J. Chem. Phys.},
  publisher = {AIP Publishing},
  author    = {Blunt,  N. S. and Booth,  George H. and Alavi,  Ali},
  year      = {2017},
  month     = {June},
  pages = {244105}
}

@article{10.1063/1.457480,
  author  = {Flyvbjerg, H. and Petersen, H. G.},
  title   = {Error estimates on averages of correlated data},
  journal = {J. Chem. Phys.},
  volume  = {91},
  number  = {1},
  pages   = {461--466},
  year    = {1989},
  month   = {07},
  issn    = {0021-9606},
  doi     = {10.1063/1.457480},
  url     = {https://doi.org/10.1063/1.457480}
}

@article{Booth2012_3,
  title     = {An explicitly correlated approach to basis set incompleteness in full configuration interaction quantum Monte Carlo},
  volume    = {137},
  issn      = {1089-7690},
  url       = {http://dx.doi.org/10.1063/1.4762445},
  doi       = {10.1063/1.4762445},
  number    = {16},
  journal   = {J. Chem. Phys.},
  publisher = {AIP Publishing},
  author    = {Booth,  George H. and Cleland,  Deidre and Alavi,  Ali and Tew,  David P.},
  year      = {2012},
  month     = {Oct},
  pages = {164112}
}

@article{Overy2014,
  title     = {Unbiased reduced density matrices and electronic properties from full configuration interaction quantum Monte Carlo},
  volume    = {141},
  issn      = {1089-7690},
  url       = {http://dx.doi.org/10.1063/1.4904313},
  doi       = {10.1063/1.4904313},
  number    = {24},
  journal   = {J. Chem. Phys.},
  publisher = {AIP Publishing},
  author    = {Overy,  Catherine and Booth,  George H. and Blunt,  N. S. and Shepherd,  James J. and Cleland,  Deidre and Alavi,  Ali},
  year      = {2014},
  month     = {Dec},
  pages = {244117}
}

@article{Guther2020,
  title     = {NECI: $N$-Electron Configuration Interaction with an emphasis on state-of-the-art stochastic methods},
  volume    = {153},
  issn      = {1089-7690},
  url       = {http://dx.doi.org/10.1063/5.0005754},
  doi       = {10.1063/5.0005754},
  number    = {3},
  journal   = {J. Chem. Phys.},
  publisher = {AIP Publishing},
  author    = {Guther,  Kai and Anderson,  Robert J. and Blunt,  Nick S. and Bogdanov,  Nikolay A. and Cleland,  Deidre and Dattani,  Nike and Dobrautz,  Werner and Ghanem,  Khaldoon and Jeszenszki,  Peter and Liebermann,  Niklas and Manni,  Giovanni Li and Lozovoi,  Alexander Y. and Luo,  Hongjun and Ma,  Dongxia and Merz,  Florian and Overy,  Catherine and Rampp,  Markus and Samanta,  Pradipta Kumar and Schwarz,  Lauretta R. and Shepherd,  James J. and Smart,  Simon D. and Vitale,  Eugenio and Weser,  Oskar and Booth,  George H. and Alavi,  Ali},
  year      = {2020},
  month     = {July},
  pages = {034107}
}

@article{PhysRevLett.118.032502,
  title     = {Nucleus-Dependent Valence-Space Approach to Nuclear Structure},
  author    = {Stroberg, S. R. and Calci, A. and Hergert, H. and Holt, J. D. and Bogner, S. K. and Roth, R. and Schwenk, A.},
  journal   = {Phys. Rev. Lett.},
  volume    = {118},
  issue     = {3},
  pages     = {032502},
  numpages  = {6},
  year      = {2017},
  month     = {Jan},
  publisher = {American Physical Society},
  doi       = {10.1103/PhysRevLett.118.032502},
  url       = {https://link.aps.org/doi/10.1103/PhysRevLett.118.032502}
}

@article{PhysRevX.10.011041,
  title         = {Direct Comparison of Many-Body Methods for Realistic Electronic Hamiltonians},
  author        = {Williams, Kiel T. and Yao, Yuan and Li, Jia and Chen, Li and Shi, Hao and Motta, Mario and Niu, Chunyao and Ray, Ushnish and Guo, Sheng and Anderson, Robert J. and Li, Junhao and Tran, Lan Nguyen and Yeh, Chia-Nan and Mussard, Bastien and Sharma, Sandeep and Bruneval, Fabien and van Schilfgaarde, Mark and Booth, George H. and Chan, Garnet Kin-Lic and Zhang, Shiwei and Gull, Emanuel and Zgid, Dominika and Millis, Andrew and Umrigar, Cyrus J. and Wagner, Lucas K.},
  collaboration = {Simons Collaboration on the Many-Electron Problem},
  journal       = {Phys. Rev. X},
  volume        = {10},
  issue         = {1},
  pages         = {011041},
  numpages      = {9},
  year          = {2020},
  month         = {Feb},
  publisher     = {American Physical Society},
  doi           = {10.1103/PhysRevX.10.011041},
  url           = {https://link.aps.org/doi/10.1103/PhysRevX.10.011041}
}

@article{Spencer2019,
  title     = {The HANDE-QMC Project: Open-Source Stochastic Quantum Chemistry from the Ground State Up},
  volume    = {15},
  issn      = {1549-9626},
  url       = {http://dx.doi.org/10.1021/acs.jctc.8b01217},
  doi       = {10.1021/acs.jctc.8b01217},
  number    = {3},
  journal   = {J. Chem. Theory Comput.},
  publisher = {American Chemical Society (ACS)},
  author    = {Spencer,  James S. and Blunt,  Nick S. and Choi,  Seonghoon and Etrych,  Ji{\v{r}}{\'\i} and Filip,  Maria-Andreea and Foulkes,  W. M. C. and Franklin,  Ruth S. T. and Handley,  Will J. and Malone,  Fionn D. and Neufeld,  Verena A. and Di Remigio,  Roberto and Rogers,  Thomas W. and Scott,  Charles J. C. and Shepherd,  James J. and Vigor,  William A. and Weston,  Joseph and Xu,  RuQing and Thom,  Alex J. W.},
  year      = {2019},
  month     = {Jan},
  pages     = {1728--1742}
}

@misc{hu2026fciqmc,
  title         = {{\textit{Ab initio}} Exact Calculation of Strongly-Correlated Nucleonic Matter},
  author        = {Rongzhe Hu and Shaoliang Jin and Xin Zhen and Haoyu Shang and Junchen Pei and Furong Xu and Francesco Marino},
  year          = {2026},
  eprint        = {2508.09252},
  archiveprefix = {arXiv},
  primaryclass  = {nucl-th},
  url           = {https://arxiv.org/abs/2508.09252}
}

@article{Noga1987,
  title     = {Towards a full CCSDT model for electron correlation. CCSDT-n models},
  volume    = {134},
  issn      = {0009-2614},
  url       = {http://dx.doi.org/10.1016/0009-2614(87)87107-5},
  doi       = {10.1016/0009-2614(87)87107-5},
  number    = {2},
  journal   = {Chem. Phys. Lett.},
  publisher = {Elsevier BV},
  author    = {Noga,  Jozef and Bartlett,  Rodney J. and Urban,  Miroslav},
  year      = {1987},
  month     = {Feb},
  pages     = {126--132}
}

@article{PhysRevLett.110.192502,
  title     = {Optimized Chiral Nucleon-Nucleon Interaction at Next-to-Next-to-Leading Order},
  author    = {Ekstr\"om, A. and Baardsen, G. and Forss\'en, C. and Hagen, G. and Hjorth-Jensen, M. and Jansen, G. R. and Machleidt, R. and Nazarewicz, W. and Papenbrock, T. and Sarich, J. and Wild, S. M.},
  journal   = {Phys. Rev. Lett.},
  volume    = {110},
  issue     = {19},
  pages     = {192502},
  numpages  = {5},
  year      = {2013},
  month     = {May},
  publisher = {American Physical Society},
  doi       = {10.1103/PhysRevLett.110.192502},
  url       = {https://link.aps.org/doi/10.1103/PhysRevLett.110.192502}
}

@article{Jin2025,
  title     = {Full configuration interaction quantum Monte Carlo in nuclear structure calculations},
  volume    = {36},
  issn      = {2210-3147},
  url       = {http://dx.doi.org/10.1007/s41365-025-01790-5},
  doi       = {10.1007/s41365-025-01790-5},
  number    = {11},
  journal   = {Nucl. Sci. Tech.},
  publisher = {Springer Science and Business Media LLC},
  author    = {Jin,  Shao-Liang and Li,  Jian-Guo and Gao,  Yuan and Hu,  Rong-Zhe and Xu,  Fu-Rong},
  year      = {2025},
  month     = {Aug},
  pages={212}
}

@article{q3vn-8y8s,
  title     = {Stochastic many-body perturbation theory for high-order calculations},
  author    = {Zhen, X. and Hu, R. Z. and Pei, J. C. and Xu, F. R.},
  journal   = {Phys. Rev. C},
  volume    = {113},
  issue     = {5},
  pages     = {L051302},
  numpages  = {7},
  year      = {2026},
  month     = {May},
  publisher = {American Physical Society},
  doi       = {10.1103/q3vn-8y8s},
  url       = {https://link.aps.org/doi/10.1103/q3vn-8y8s}
}

@article{Machleidt2023,
  title     = {What is {\textit{ab initio}}?},
  volume    = {64},
  issn      = {1432-5411},
  url       = {http://dx.doi.org/10.1007/s00601-023-01857-2},
  doi       = {10.1007/s00601-023-01857-2},
  number    = {4},
  pages={77},
  journal   = {Few-Body Syst.},
  publisher = {Springer Science and Business Media LLC},
  author    = {Machleidt,  R.},
  year      = {2023},
  month     = {Oct}
}

@article{Ekstrm2023,
  title     = {What is {\textit{ab initio}} in nuclear theory?},
  volume    = {11},
  issn      = {2296-424X},
  pages={1129094},
  url       = {http://dx.doi.org/10.3389/fphy.2023.1129094},
  doi       = {10.3389/fphy.2023.1129094},
  journal   = {Front. Phys.},
  publisher = {Frontiers Media SA},
  author    = {Ekstr\"{o}m,  A. and Forss\'en,  C. and Hagen,  G. and Jansen,  G. R. and Jiang,  W. and Papenbrock,  T.},
  year      = {2023},
  month     = {Feb}
}

@article{Walker1977,
  title     = {An Efficient Method for Generating Discrete Random Variables with General Distributions},
  volume    = {3},
  issn      = {1557-7295},
  url       = {http://dx.doi.org/10.1145/355744.355749},
  doi       = {10.1145/355744.355749},
  number    = {3},
  journal   = {ACM Trans. Math. Softw.},
  publisher = {Association for Computing Machinery (ACM)},
  author    = {Walker,  Alastair J.},
  year      = {1977},
  month     = {Sept},
  pages     = {253--256}
}

@article{Holmes2016,
  title     = {Heat-Bath Configuration Interaction: An Efficient Selected Configuration Interaction Algorithm Inspired by Heat-Bath Sampling},
  volume    = {12},
  issn      = {1549-9626},
  url       = {http://dx.doi.org/10.1021/acs.jctc.6b00407},
  doi       = {10.1021/acs.jctc.6b00407},
  number    = {8},
  journal   = {J. Chem. Theory Comput.},
  publisher = {American Chemical Society (ACS)},
  author    = {Holmes,  Adam A. and Tubman,  Norm M. and Umrigar,  C. J.},
  year      = {2016},
  month     = {Aug},
  pages     = {3674--3680}
}

@article{Holmes2016_2,
  title     = {Efficient Heat-Bath Sampling in Fock Space},
  volume    = {12},
  issn      = {1549-9626},
  url       = {http://dx.doi.org/10.1021/acs.jctc.5b01170},
  doi       = {10.1021/acs.jctc.5b01170},
  number    = {4},
  journal   = {J. Chem. Theory Comput.},
  publisher = {American Chemical Society (ACS)},
  author    = {Holmes,  Adam A. and Changlani,  Hitesh J. and Umrigar,  C. J.},
  year      = {2016},
  month     = {Mar},
  pages     = {1561--1571}
}

@article{PhysRevC.102.051303,
  title     = {Charge radii of exotic neon and magnesium isotopes},
  author    = {Novario, S. J. and Hagen, G. and Jansen, G. R. and Papenbrock, T.},
  journal   = {Phys. Rev. C},
  volume    = {102},
  issue     = {5},
  pages     = {051303(R)},
  numpages  = {8},
  year      = {2020},
  month     = {Nov},
  publisher = {American Physical Society},
  doi       = {10.1103/PhysRevC.102.051303},
  url       = {https://link.aps.org/doi/10.1103/PhysRevC.102.051303}
}

@article{Hagen2015,
  title     = {Neutron and weak-charge distributions of the $^{48}\mathrm{Ca}$ nucleus},
  volume    = {12},
  issn      = {1745-2481},
  url       = {http://dx.doi.org/10.1038/nphys3529},
  doi       = {10.1038/nphys3529},
  number    = {2},
  journal   = {Nat. Phys.},
  publisher = {Springer Science and Business Media LLC},
  author    = {Hagen,  G. and Ekstr\"{o}m,  A. and Forss\'en,  C. and Jansen,  G. R. and Nazarewicz,  W. and Papenbrock,  T. and Wendt,  K. A. and Bacca,  S. and Barnea,  N. and Carlsson,  B. and Drischler,  C. and Hebeler,  K. and Hjorth-Jensen,  M. and Miorelli,  M. and Orlandini,  G. and Schwenk,  A. and Simonis,  J.},
  year      = {2015},
  month     = {Nov},
  pages     = {186--190}
}

@article{10.1093/ptep/ptac097,
  author        = {Workman, R. L. and others},
  collaboration = {Particle Data Group},
  title         = {Review of Particle Physics},
  journal       = {Prog. Theor. Exp. Phys.},
  volume        = {2022},
  number        = {8},
  pages         = {083C01},
  year          = {2022},
  month         = {08},
  issn          = {2050-3911},
  doi           = {10.1093/ptep/ptac097},
  url           = {https://doi.org/10.1093/ptep/ptac097}
}

@article{Mller2025,
  title     = {The nuclear charge radius of $^{13}\mathrm{C}$},
  volume    = {16},
  issn      = {2041-1723},
  url       = {http://dx.doi.org/10.1038/s41467-025-60280-9},
  doi       = {10.1038/s41467-025-60280-9},
  number    = {1},
  pages={6234},
  journal   = {Nat. Commun.},
  publisher = {Springer Science and Business Media LLC},
  author    = {M\"{u}ller,  Patrick and Heinz,  Matthias and Imgram,  Phillip and K\"{o}nig,  Kristian and Maass,  Bernhard and Miyagi,  Takayuki and N\"{o}rtersh\"{a}user,  Wilfried and Roth,  Robert and Schwenk,  Achim},
  year      = {2025},
  month     = {July}
}

@article{PhysRevC.102.054301,
  title     = {Accurate bulk properties of nuclei from $A=2$ to $\ensuremath{\infty}$ from potentials with $\mathrm{\ensuremath{\Delta}}$ isobars},
  author    = {Jiang, W. G. and Ekstr\"om, A. and Forss\'en, C. and Hagen, G. and Jansen, G. R. and Papenbrock, T.},
  journal   = {Phys. Rev. C},
  volume    = {102},
  issue     = {5},
  pages     = {054301},
  numpages  = {8},
  year      = {2020},
  month     = {Nov},
  publisher = {American Physical Society},
  doi       = {10.1103/PhysRevC.102.054301},
  url       = {https://link.aps.org/doi/10.1103/PhysRevC.102.054301}
}

@article{PhysRevC.80.021306,
  title     = {{\textit{Ab initio}} computation of neutron-rich oxygen isotopes},
  author    = {Hagen, G. and Papenbrock, T. and Dean, D. J. and Hjorth-Jensen, M. and Asokan, B. Velamur},
  journal   = {Phys. Rev. C},
  volume    = {80},
  issue     = {2},
  pages     = {021306(R)},
  numpages  = {5},
  year      = {2009},
  month     = {Aug},
  publisher = {American Physical Society},
  doi       = {10.1103/PhysRevC.80.021306},
  url       = {https://link.aps.org/doi/10.1103/PhysRevC.80.021306}
}

@article{Cleland2010,
  title = {Communications: Survival of the fittest: Accelerating convergence in full configuration-interaction quantum Monte Carlo},
  volume = {132},
  ISSN = {1089-7690},
  url = {http://dx.doi.org/10.1063/1.3302277},
  DOI = {10.1063/1.3302277},
  number = {4},
  journal = {J. Chem. Phys.},
  publisher = {AIP Publishing},
  author = {Cleland,  Deidre and Booth,  George H. and Alavi,  Ali},
  year = {2010},
  month = {Jan},
  pages = {041103}
}

@article{Gloeckner1974,
  title     = {Spurious center-of-mass motion},
  volume    = {53},
  issn      = {0370-2693},
  url       = {http://dx.doi.org/10.1016/0370-2693(74)90390-6},
  doi       = {10.1016/0370-2693(74)90390-6},
  number    = {4},
  journal   = {Phys. Lett. B},
  publisher = {Elsevier BV},
  author    = {Gloeckner,  D. H. and Lawson,  R. D.},
  year      = {1974},
  month     = {Dec},
  pages     = {313--318}
}

@article{PhysRevC.102.034320,
  title     = {{\textit{Ab initio}} multishell valence-space Hamiltonians and the island of inversion},
  author    = {Miyagi, T. and Stroberg, S. R. and Holt, J. D. and Shimizu, N.},
  journal   = {Phys. Rev. C},
  volume    = {102},
  issue     = {3},
  pages     = {034320},
  numpages  = {10},
  year      = {2020},
  month     = {Sep},
  publisher = {American Physical Society},
  doi       = {10.1103/PhysRevC.102.034320},
  url       = {https://link.aps.org/doi/10.1103/PhysRevC.102.034320}
}

@article{Jiang2026,
  title = {Toward ab initio quantum simulations of atomic nuclei using noisy qubits},
  volume = {71},
  ISSN = {2095-9273},
  url = {http://dx.doi.org/10.1016/j.scib.2026.02.052},
  DOI = {10.1016/j.scib.2026.02.052},
  number = {7},
  journal = {Science Bulletin},
  publisher = {Elsevier BV},
  author = {Jiang,  Chongji and Pei,  Junchen and Hu,  Rongzhe and Jin,  Shaoliang and Shang,  Haoyu and Fan,  Siqin and Xu,  Furong},
  year = {2026},
  month = {Apr},
  pages = {1598–1601}
}

@article{gjmd-fyjy,
  title = {Chiral three-nucleon forces for the new local position-space two-nucleon potential in \textit{ab initio} many-body calculations},
  author = {Hu, R. Z. and Li, J. G. and Fan, S. Q. and Xu, F. R.},
  journal = {Phys. Rev. C},
  pages = {},
  year = {2026},
  month = {Jun},
  publisher = {American Physical Society},
  doi = {10.1103/gjmd-fyjy},
  url = {https://link.aps.org/doi/10.1103/gjmd-fyjy}
}

\end{document}